\documentclass[letterpaper,10pt]{article}
\usepackage[totalwidth=480pt,totalheight=680pt]{geometry}
\usepackage{xcolor}

\usepackage[textsize=scriptsize]{todonotes}

\AddToHook{cmd/@makefntext/before}{}

\usepackage{microtype}
\usepackage{amsmath,amssymb,amsthm}
\usepackage{enumerate}
\usepackage[authoryear,square,comma]{natbib}
\usepackage{booktabs}
\usepackage{graphicx}
\usepackage{placeins}
\usepackage{float}

\usepackage{needspace}
\usepackage{aliascnt}
\floatstyle{ruled}
\newfloat{algorithm}{tbp}{loa}
\floatname{algorithm}{Algorithm}
\floatstyle{plain}
\definecolor{darkgreen}{rgb}{0,0.5,0}
\usepackage{hyperref}

\hypersetup{
  colorlinks=true,
  linkcolor=red,
  citecolor=darkgreen,
  urlcolor=blue,
  filecolor=magenta,
  linktoc=all,
  pdftitle={Mechanism Design for Public Health Referrals}
}
\usepackage[capitalize,nameinlink,noabbrev]{cleveref}
\usepackage{etoc}
\crefname{algorithm}{Algorithm}{Algorithms}
\Crefname{algorithm}{Algorithm}{Algorithms}

\newtheorem{theorem}{Theorem}[section]
\newaliascnt{corollary}{theorem}

\aliascntresetthe{corollary}
\crefname{corollary}{Corollary}{Corollaries}
\Crefname{corollary}{Corollary}{Corollaries}

\newaliascnt{proposition}{theorem}
\newtheorem{proposition}[proposition]{Proposition}
\aliascntresetthe{proposition}
\crefname{proposition}{Proposition}{Propositions}
\Crefname{proposition}{Proposition}{Propositions}

\newaliascnt{lemma}{theorem}
\newtheorem{lemma}[lemma]{Lemma}
\aliascntresetthe{lemma}
\crefname{lemma}{Lemma}{Lemmas}
\Crefname{lemma}{Lemma}{Lemmas}

\newcommand{\restatedtitle}{}
\newtheorem*{restatedresult}{\restatedtitle}
\newenvironment{restatement}[1]
  {\Needspace{8\baselineskip}\renewcommand{\restatedtitle}{#1}\begin{restatedresult}}
  {\end{restatedresult}}
\theoremstyle{definition}
\newaliascnt{definition}{theorem}
\newtheorem{definition}[definition]{Definition}
\aliascntresetthe{definition}
\crefname{definition}{Definition}{Definitions}
\Crefname{definition}{Definition}{Definitions}

\newaliascnt{example}{theorem}
\newtheorem{example}[example]{Example}
\aliascntresetthe{example}
\crefname{example}{Example}{Examples}
\Crefname{example}{Example}{Examples}

\theoremstyle{remark}
\newaliascnt{remark}{theorem}
\newtheorem{remark}[remark]{Remark}
\aliascntresetthe{remark}
\crefname{remark}{Remark}{Remarks}
\Crefname{remark}{Remark}{Remarks}

\newtheorem*{remark*}{Remark}

\DeclareMathOperator*{\argmax}{arg\,max}

\numberwithin{equation}{section}
\allowdisplaybreaks

\title{Mechanism Design for Public Health Referrals}
\makeatletter
\def\@fnsymbol#1{\ensuremath{\ifcase#1\or \ast\or \dagger\or \ddagger\or
  \mathsection\or \mathparagraph\or \|\or \ast\ast\or \dagger\dagger
  \or \ddagger\ddagger \else\@ctrerr\fi}}
\makeatother
\author{%
Tzeh Yuan Neoh\textsuperscript{1,$\ast$}\qquad
Davin Choo\textsuperscript{1}\qquad
Melissa Dutz\textsuperscript{2}\qquad
Yuqi Pan\textsuperscript{1}\\[0.6ex]
Avrim Blum\textsuperscript{2}\qquad
\'Eva Tardos\textsuperscript{3}\qquad
Milind Tambe\textsuperscript{1}\\[2.2ex]
{\normalsize \textsuperscript{1}Harvard University\qquad
\textsuperscript{2}Toyota Technological Institute at Chicago\qquad
\textsuperscript{3}Cornell University}\\[0.8ex]
{\small \textsuperscript{$\ast$}Corresponding author:
\href{mailto:tzehyuan_neoh@g.harvard.edu}{\texttt{tzehyuan\_neoh@g.harvard.edu}}}%
}
\date{}

\begin{document}
\maketitle

\begin{abstract}
Public-health referral programmes recruit through participants' social
networks. Motivated by HIV self-testing, we study how to maximise
recruitment when each participant privately knows the distribution of
how many contacts they can recruit, while the programme knows only a
population prior over these distributions.  The programme offers a
fixed menu of contracts, each specifying a number of vouchers and a
payment rule, and must work within two budgets, one for vouchers and
one for rewards. Recruitment-maximising
menus can be complex and are strongly NP-hard to compute, and in
general no fixed menu guarantees a constant fraction of the
recruitment achieved by an oracle that observes participants'
distributions. We identify two conditions under which simple
mechanisms narrow this gap. When the reward budget is large relative
to the voucher budget, a voucher-price mechanism
asymptotically matches the oracle. When participants' distributions
are log-concave, a completion-bonus mechanism
asymptotically achieves a \(1-1/e\)
fraction of the oracle's recruitment even with equal budgets, and this
guarantee is tight.
Simulations based on HIV transmission network data illustrate
the mechanisms' complementary strengths and show that both can
outperform a constant-allocation baseline motivated by common practice.
\end{abstract}

\etocsettagdepth{mainmatter}{section}
\etocsettagdepth{appendix}{none}
\tableofcontents
\newpage

\etocdepthtag.toc{mainmatter}
\section{Introduction}
\label{sec:introduction}

Network interventions in public health leverage social relationships to disseminate health information, promote behavioural change, and recruit individuals into health programmes \citep{valente2012network}. Recognising that individuals differ in their ability to reach and influence others, a growing body of work uses artificial intelligence and algorithmic methods to identify effective individuals for public-health interventions. Examples include selecting influential peer leaders for HIV prevention \citep{yadav2016using,wilder2021clinical}, identifying individuals who can promote vaccination uptake \citep{fugenschuh2023overcoming,neophytou2024promoting}, and targeting individuals to encourage the adoption of preventive health practices \citep{alexander2022algorithms}.

These approaches typically predict how well individuals can reach others from recorded data. However, machine-learning methods in public health are often limited by the quantity and quality of the available data \citep{pinto2025machine}, and suitable datasets remain scarce in many low- and middle-income countries \citep{ciecierskiholmes2022artificial,alami2020artificial}. Even with abundant data, recorded characteristics may omit factors such as how trusted a participant is in their community or their contacts' willingness to participate. Participants may therefore know their referral prospects better than predictions based on observable characteristics would suggest.

At the same time, there has been an increasing emphasis on individual autonomy and participation in healthcare decisions \citep{who2021ethics}. In the foreword to the WHO Guideline on Self-Care Interventions for Health and Well-Being, the WHO Director-General advocates a people-centred approach that views individuals as ``active decision-makers in their own health, not merely passive recipients of health services'' \citep{ghebreyesus2022foreword}. Together, these considerations motivate giving participants a choice over how they contribute to public-health network interventions, drawing on their own knowledge of their social connections and their ability to reach others.

\paragraph{Our Setting: Network Recruitment with Self-Testing.}
Our motivating setting is network recruitment with HIV self-testing,
in which participants distribute test kits to people in their social
networks and receive rewards for successful referrals. Monetary
incentives can expand the reach of such programmes:
\citet{zhou2022monetary} show that, compared with distributing kits
without monetary incentives, adding rewards for verified tests
increases the number of contacts recruited to test, including
first-time testers.

Standard practice in respondent-driven recruitment, from its
introduction \citep{heckathorn1997respondent} to WHO guidance \citep{who2013rds}, is to give each participant
a small, fixed number of referral coupons, commonly three
\citep{goel2009respondent}.
Participants, however, differ in how many people they can recruit:
some leave part of their allocation unused, while others could reach
more people with additional resources. This motivates adapting
allocations to participants' referral prospects to increase recruitment. Recent algorithmic
approaches use known or predicted referral-capacity distributions to
decide how many resources each participant should receive
\citep{pan2026adaptive,kong2026generative}. We address the same
allocation question but ask how participants' \emph{own} knowledge of
their referral prospects can guide the distribution of limited
resources.

\paragraph{What Does a Mechanism-Design Approach Entail?}
The programme has a budget of \(B_1\) vouchers and a
budget of \(B_2\) reward units, and aims to maximise expected
recruitment; there is a \emph{reward surplus} when \(B_2>B_1\). Each
participant privately knows the distribution of the number of contacts
they could recruit, which we call their \emph{type}, whereas the
programme knows only a population prior over types. It offers a menu of contracts, each pairing a number
of vouchers with a payout rule that specifies the reward for each
number of successful referrals, and each participant chooses the
contract that maximises their expected reward. Because referrals and
payments occur later, a chosen contract consumes its vouchers and
earmarks its maximum payout; participants are reached until a budget
can no longer cover the menu. To illustrate, we consider two basic mechanisms in which each
participant receives exactly the number of vouchers they request; the
mechanisms differ only in their payout rules.

\begin{example}[Linear payments]
\label{ex:intro-linear}
Each successful referral earns one unit of reward. An additional
voucher pays if it is used and costs the participant nothing
otherwise, so participants request every voucher that has any chance
of being used. Consider a participant who recruits \(2\) contacts with
probability \(0.9\) and \(8\) with probability \(0.1\). This
participant requests \(8\) vouchers, and the programme earmarks \(8\)
units of reward, although expected recruitment is only \(2.6\):
vouchers \(3\) to \(8\) are each used with probability \(0.1\). Linear payments thus encourage
over-requesting: with limited budgets, these vouchers and their
earmarked rewards would recruit more if given to participants certain
to use them.
\end{example}

\begin{example}[All-or-nothing payments]
\label{ex:intro-all-or-nothing}
A participant holding \(q\) vouchers earns \(q\) units of reward only
if all \(q\) referrals succeed. Consider a participant
who recruits \(5\) contacts with probability \(0.2\) and \(6\) with
probability \(0.8\). Requesting \(5\) vouchers guarantees a reward of
\(5\), whereas \(6\) vouchers yield an expected reward of only
\(6\times0.8=4.8\), so the participant requests \(5\), although voucher
\(6\) would be used with probability \(0.8\). All-or-nothing payments thus encourage
under-requesting: a voucher that would be used with probability \(0.8\)
goes unallocated, losing recruitment that the budgets could have
supported.
\end{example}

These examples raise the central question of this paper: given a
population prior and two budgets, how should we design simple
mechanisms that recruit well? We make this question precise in
\cref{sec:contributions}.

\subsection{Our Contributions}
\label{sec:contributions}

To our knowledge, this is the first application of mechanism design to
public-health referral recruitment: instead of predicting participants' referral prospects, the programme lets participants reveal them through the contracts they choose. Our main technical results are two families of simple mechanisms, voucher-price and completion-bonus
mechanisms, that work well in different regimes. The voucher-price mechanism performs
well when the reward budget is sufficiently large relative to the
voucher budget. The
completion-bonus mechanism needs no such surplus but relies on
log-concave referral distributions: it asymptotically attains a
\(1-1/e\approx0.632\) fraction of the oracle's recruitment, which is
tight. We complement them with an impossibility result:
without either restrictions on referral distributions or a reward
surplus, no
fixed menu can guarantee more than an \(O(1/\sqrt m)\) fraction of the
oracle's recruitment, where \(m\) is the maximum number of
contacts a participant can recruit. We also discuss the limitations of recruitment-maximising
menus and evaluate both mechanisms in simulations
on a population prior built from an ICPSR HIV dataset.

\begin{itemize}
\setlength{\itemsep}{3pt}
\item \textbf{Model (\cref{sec:preliminaries}).} We
formulate the joint allocation of vouchers and referral rewards when
each participant's referral capacity, the number of contacts they can
recruit, is an integer between \(0\) and a known maximum \(m\), and
participants privately know the distribution of their own capacity.
The designer knows a population prior and offers a fixed menu of
contracts, each pairing a number
of vouchers with a payout rule. Each chosen contract commits its
vouchers and reserves its maximum payout before any referrals occur. We introduce an expected-budget
relaxation that is easier to analyse. For bounded payouts, it matches each menu's sequential recruitment up to a
\(1-o(1)\) factor as the effective budget grows
(\cref{thm:relaxation-transfer}).

\item \textbf{Recruitment-maximising mechanisms
(\cref{sec:recruitment-maximizing}).} We identify three limitations of recruitment-maximising
menus. First, optimal menus can be difficult
for participants to reason about: even a small instance requires more
contracts than there are possible outcomes (\cref{ex:four-contracts}). Second, optimal menus are
computationally hard to find: the problem is strongly NP-hard even for
\(m=9\) and finitely many types (\cref{thm:menu-hardness}). Third,
even an optimal menu can
recruit far fewer people than an allocation that uses type information
directly; we quantify this gap in \cref{sec:general}
(\cref{prop:warmup-impossibility}). When this gap is large, a
predictive approach, for example one that estimates referral prospects
from participant characteristics, may be preferable. Choosing between
the approaches therefore requires a benchmark that measures this gap.
The first two limitations
motivate the simple, efficiently computable menus that we study; the
third motivates the following benchmark.

\item \textbf{Oracle benchmark (\cref{sec:oracle-benchmark}).} We
define an oracle that observes each participant's type, but not their
realised capacity, and assigns vouchers without incentive constraints.
We show that a threshold rule, which gives each participant one voucher plus every additional
voucher whose probability of being used exceeds a cutoff shared by all
participants, is optimal for the expected-budget relaxation and
asymptotically optimal for sequential recruitment. We provide an
algorithm that computes the cutoff and its tie-breaking rule
(\cref{alg:oracle-threshold,thm:simple-threshold}). This benchmark
serves as the basis for evaluating the mechanisms that follow.

\item \textbf{General types (\cref{sec:general}).} We show that, for every population, a
\emph{voucher-price} mechanism reproduces the allocation of the
optimal threshold rule exactly. It therefore attains the oracle's
expected recruitment per voucher. However, the stronger incentives it
offers require larger earmarked rewards, so the mechanism may reach
fewer participants. Its approximation factor,
\(\min\{1,\max\{1,B_2/B_1\}/(1+\theta^\star(m-1))\}-o(1)\), depends on
the optimal threshold \(\theta^\star\) and on the budget ratio
\(B_2/B_1\) (\cref{thm:threshold-reward-guarantee}). The mechanism
thus performs well in regimes where \(B_2/B_1\) is high or
\(\theta^\star\) is low. Without reward surplus, however, no
admissible fixed menu can guarantee more than an \(O(1/\sqrt m)\)
fraction of the oracle's recruitment
(\cref{prop:warmup-impossibility}). This impossibility shows why
distributional structure matters: guarantees independent of \(m\)
require restricting the referral distributions.

\item \textbf{Log-concave types (\cref{sec:restricted}).} The
\(O(1/\sqrt m)\) bound is driven by participants whose capacity is
either \(0\) or \(m\). We therefore ask whether stronger guarantees
hold for \emph{log-concave} types, whose probabilities satisfy
\(D(j)^2\ge D(j-1)D(j+1)\) on an integer interval; this family
includes uniform, binomial, and truncated geometric distributions. We
define mechanisms that combine linear payments with a completion bonus
and provide an algorithm that computes one (\cref{alg:restricted-menu}).
We show that these mechanisms achieve a \(1-1/e\) fraction of the
oracle's recruitment for log-concave types, up to \(o(1)\) terms
(\cref{thm:restricted-guarantees}). Conversely, no admissible fixed
menu can guarantee a worst-case ratio above \(1-1/e\)
(\cref{thm:uniform-upper}), so the guarantee is tight.
\item \textbf{Experiments (\cref{sec:experiments}).} On a population
prior built from HIV transmission network data, both mechanisms
outperform a constant allocation that ignores participants'
information, and the parameters chosen by our algorithms are close to
the best parameters found by tuning. Truncating \(m\) in the
voucher-price menu can substantially increase its recruitment by
reducing the rewards it must reserve. The voucher price outperforms the
completion bonus when the reward budget is sufficiently large relative
to the voucher budget, whereas the completion bonus outperforms the
voucher price when \(B_2=B_1\). When capacity distributions become
more extreme, the completion bonus degrades, while the voucher price
declines only mildly. Interestingly, under a misspecified prior, the
two mechanisms are robust in complementary ways: the completion bonus holds up when
the prior overestimates recruitment, and the voucher price with a
sufficiently conservative cap holds up when it underestimates
recruitment. In each case, the mechanism can outperform the oracle
threshold policy fitted on the same wrong prior.
\end{itemize}

\subsection{Related Work}
\label{sec:related-work}

\paragraph{Bayesian mechanism design and information elicitation.}
Bayesian mechanism design studies allocation when agents hold private
types drawn from a known prior \citep{myerson1981optimal}. For welfare
maximisation with independent single-parameter types,
\citet{hartline2010bayesian} show how to convert approximation
algorithms into Bayesian incentive-compatible mechanisms with
essentially the same approximation guarantees, and simple posted-price
mechanisms approximate optimal revenue in several multi-parameter
settings \citep{chawla2010posted}. Related work studies payments under
hard budgets \citep{singer2010budget,goel2025budget}, simple and
menu-based outcome-contingent contracts
\citep{duetting2019simple,castiglioni2022menus,guruganesh2023menus},
and allocation when agents privately know their probability of
success \citep{porter2008fault}. A complementary literature elicits
private information directly. Proper scoring rules reward
probabilistic forecasts against observed outcomes
\citep{gneiting2007proper}; properties such as quantiles can be
elicited without the full distribution \citep{lambert2009eliciting};
and recent work designs bounded scoring rules to incentivise costly
effort or information acquisition
\citep{hartline2023effort,chen2023learning}. When only the outcome of
the selected action is observed, as in decision markets
\citep{chen2011decision}, or when forecasters care about the decisions
their reports induce \citep{boutilier2012eliciting}, incentives must
account for this dependence.

Our main departure from both literatures is that incentives are
coupled with reach: every selected contract consumes vouchers and
reserves its maximum payout from a reward budget, so the rewards that
screen participants' referral prospects also determine how many
participants the programme can reach. Elicitation is therefore costly.
Under the voucher-price menu of \cref{sec:easier-regimes}, for example, each participant faces a
newsvendor problem and requests a \((1-\theta^\star)\)-quantile of
their capacity distribution, exactly the statistic used by the
oracle's threshold rule. Inducing this choice, however, reserves an
extra \(\theta^\star(m-q)\) in rewards beyond the \(q\) that the
oracle reserves, so when the reward budget binds, fewer participants
can be reached. Unlike in the
welfare settings above, matching a type-aware allocation can thus
carry a substantial incentive cost. Our setting also shares, in a
specific form, the dependence of feedback on decisions: giving \(q\)
vouchers reveals only \(\min\{R,q\}\), so capacities at or above
\(q\) are indistinguishable, and participants' preferences over
quantities arise entirely through expected rewards.

\paragraph{AI for public-health resource allocation.}
AI methods increasingly guide the allocation of scarce public-health
resources. \citet{wilder2021clinical} evaluate AI-selected peer leaders
in a clinical trial of an HIV-prevention intervention for youth
experiencing homelessness, while \citet{mate2022field} deploy
restless-bandit policies to prioritise health-worker calls in a
maternal and child health programme.
\citet{kangaslahti2026policy} study sequential HIV testing
on transmission networks that are revealed as testing proceeds,
embedding a generative model of how the network expands into the
testing policy.
Closest to our setting, \citet{pan2026adaptive} allocate limited
referral resources across multiple rounds to participants with
stochastic referral capacities. They characterise optimal
single-round allocation through marginal survival probabilities
and develop a population-level surrogate for multi-round
planning. \citet{kong2026generative} learn models of both referral
capacity and recruit covariates, allowing allocation decisions
to account for the characteristics of future recruits.
These approaches allocate using known or predicted capacity
distributions. Our oracle benchmark applies the same marginal-gain
logic to a known type distribution; we instead design incentives that
draw on participants' private knowledge of their own referral
prospects.

\paragraph{Referral incentives.}
Referral incentives have been studied in query propagation
\citep{kleinberg2005query}, multi-level marketing
\citep{emek2011mechanisms}, and truthful influence maximisation with
privately known participation costs \citep{singer2012influence}.
Closest to our work, \citet{lobel2017customer} study how a firm should
reward customers who buy its product and refer friends. In both
settings, the designer pays for successful referrals made by
individuals who differ in how many referrals they can generate. The
settings differ in three ways. Their firm maximises profit, trading
payments against revenue, whereas our programme maximises recruitment
subject to hard voucher and reward budgets. Their rewards motivate
customers to refer, whereas ours screen participants whose referral
capacities are fixed. Finally, their firm has no limited stock to
allocate, whereas each voucher given to one participant, used or not,
is unavailable to others.
\section{Preliminaries}
\label{sec:preliminaries}

This section formalises the model. We describe the individuals and
budgets (\cref{sec:individuals}), the contracts a mechanism may offer
(\cref{sec:mechanisms}), and the sequential recruitment process
together with a relaxation that simplifies its analysis
(\cref{sec:recruitment-process}). We write \(x_+:=\max\{x,0\}\). For a set \(S\), let
\(\Delta(S)\) denote the probability distributions on \(S\), and
let \(\delta_r\) denote the point mass at \(r\).

\subsection{Individuals and Budgets}
\label{sec:individuals}

From here on, we call the programme the \emph{designer} and its
participants \emph{individuals}.
Let \(m\ge1\) bound an individual's referral capacity, so that
capacities take values in \(\Omega=\{0,\ldots,m\}\). We study one
recruitment round with \(n\ge1\) initial individuals, processed in a
fixed order. The
designer has a budget of \(B_1\in\mathbb Z_{\ge0}\) vouchers and a
reward budget \(B_2\in[0,\infty]\).

We study allocation conditional on participation. The individuals we
consider have already been engaged as referrers, and a referrer
without a voucher cannot refer anyone, so engaging an individual
commits the programme to providing at least one voucher. The available quantities are therefore
\(\mathcal Q=\{1,\ldots,m\}\), and the mechanism decides how many
additional vouchers to give and how to reward referrals.

The designer knows a population distribution
\(\mathfrak D\in\Delta(\Delta(\Omega))\), from which types are drawn
independently: \(D_i\overset{\mathrm{i.i.d.}}{\sim}\mathfrak D\).
Individual \(i\) privately knows their type \(D_i\) when choosing a
contract; only afterwards is their referral capacity drawn as
\(R_i\sim D_i\). Conditional on the types, capacities are drawn
independently across individuals and are unaffected by the chosen
contracts. An individual given \(q\) vouchers makes
\(\min\{R_i,q\}\) successful referrals.

For a type \(D\), write \(p_k(D):=\Pr_{R\sim D}(R\ge k)\)
for the probability that voucher \(k\) is used, which is also its
marginal recruitment gain. Expected recruitment from \(q\) vouchers is
\(v_D(q):=\mathbb E_{R\sim D}[\min\{R,q\}]=\sum_{k=1}^q p_k(D)\).
Finally, let \(b_+(D):=\max\{r:D(r)>0\}\) be the largest capacity in
the support of \(D\), and \(\bar b(D):=\max\{1,b_+(D)\}\); no voucher
beyond \(\bar b(D)\) is ever used.

The analysis allows arbitrary population priors. For computation, a
finite prior is specified by its types and their probabilities, all
rational. The algorithms below use this representation; their formulas
in terms of population expectations also define the mechanisms for
general priors.

\subsection{Contracts and Menus}
\label{sec:mechanisms}

We next specify what a mechanism can offer an individual. A contract \(C=(q,\tau)\) consists of a quantity \(q\in\mathcal Q\)
and a nonnegative payment rule \(\tau:\Omega\to\mathbb R_{\ge0}\).

\begin{definition}[Admissible contracts]
\label{def:admissible-contracts}
A contract \(C=(q,\tau)\) is admissible if
(i) \(\tau\) is nondecreasing;
(ii) \(\tau(z)=\tau(q)\) for \(z\ge q\); and
(iii) \(\tau(q)\ge q\).
\end{definition}

Conditions (i) and (ii) make rewards nondecreasing and dependent only
on the referrals the contract can observe. Condition (iii) reflects a
programme commitment: a participant who uses all \(q\) vouchers
receives at least a minimum reward per voucher, whose value we
normalise to one, so completing all \(q\) referrals pays at least
\(q\). This floor is what makes the reward budget meaningful: without
it, multiplying every payment in a menu by the same small factor would
leave all choices unchanged while making the earmarked rewards
negligible. It also sets a common reward scale across contracts and
menus.
Write \(\overline\tau_C:=\tau(q)\) for the maximum reward; it is
earmarked when the contract is selected.

\begin{definition}[Menu]
\label{def:menu}
A \emph{menu} is a nonempty collection of admissible contracts
\(\mathcal M=\{C_\ell=(q_\ell,\tau_\ell):\ell\in I\}\) with uniformly
bounded maximum payments.
\end{definition}

Individuals are risk-neutral, with outside-option utility zero. Under
a menu-based mechanism, type \(D\) chooses a contract maximising
their expected payment
\(
 U(C;D):=\mathbb E_{R\sim D}[\tau(\min\{R,q\})].
\)
We assume \emph{favourable tie-breaking}: the designer specifies a
measurable selection from each type's utility-maximising contracts,
possibly randomised.\footnote{Favourable tie-breaking lets us state
results for arbitrary priors, including priors with finitely many
types, under which a positive share of the population may be
indifferent between two contracts. It plays no role when ties have
probability zero: if almost every type has a unique best response in
the offered menu, as one expects when types vary continuously across
the population, then every selection rule induces the same outcome.}
The selection is fixed before recruitment and
depends only on the individual's type; randomisation is independent
across individuals and of capacities. We consider menus that admit
such a selection, as every finite menu does. The notation
\(\mathcal M\) includes the selection rule.

By the classical \emph{taxation principle} \citep{rochet1985taxation},
every deterministic individually rational (IR) and
incentive-compatible (IC) direct-revelation mechanism over our
contracts can be implemented by a menu, and every menu with its
selection rule induces a possibly randomised IR and IC
direct-revelation mechanism; \cref{app:elicitation} proves
this version for our contract class.

\subsection{Recruitment Process and Expected-Budget Relaxation}
\label{sec:recruitment-process}
\label{sec:menu-relaxation}

\paragraph{Sequential recruitment.}
Given a menu, recruitment proceeds as follows. The menu and its
selection rule are fixed throughout recruitment, so every participant
in the round faces the same available terms, regardless of when they
are reached.
Individuals are processed in turn, starting from remaining budgets
$b_1=B_1$ (vouchers) and $b_2=B_2$ (rewards). The next individual is
reached only if every offered contract can still be honoured, that is,
if $b_1\ge\sup_{\ell\in I}q_\ell$ and
$b_2\ge\sup_{\ell\in I}\overline\tau_\ell$.

A reached individual $i$ selects a best-response contract
$C_{\ell_i}$. The designer allocates $q_{\ell_i}$ vouchers and
earmarks $\overline\tau_{\ell_i}$ in rewards, deducting these amounts
from $b_1$ and $b_2$. The individual generates
$\min\{R_i,q_{\ell_i}\}$ referrals and is paid $\tau_{\ell_i}$ of that
number. Payments are settled only after the allocation phase, so
unused earmarks cannot fund later allocations in the same round.

Recruitment stops once all $n$ individuals have been processed or a
remaining budget can no longer cover the whole menu. Let $\operatorname{Rec}(\mathcal M;B_1,B_2)$ denote the
expected number of successful referrals generated by this process.

Throughout, $B:=\lfloor\min\{B_1,B_2\}\rfloor$ denotes the
\emph{effective budget}. No mechanism can allocate more than $B$
vouchers, since allocating $q$ vouchers also earmarks at least $q$
reward units by condition~(iii) of \cref{def:admissible-contracts}. Unless stated otherwise, $o(1)$
refers to $B\to\infty$ with $m$ fixed.

\paragraph{Expected-budget relaxation.}
The number of individuals reached depends on the random choices of
earlier individuals, which makes \(\operatorname{Rec}\) difficult to
analyse directly. We therefore replace the budget constraints by their
expectations. For a single individual drawn from the population, let $Q$ and $P$ be
the quantity and maximum payment of the selected contract. A fixed
menu $\mathcal M$ then has mean voucher use $u(\mathcal M):=\mathbb E[Q]$,
mean recruitment $r(\mathcal M):=\mathbb E[\min\{R,Q\}]$, and mean
earmarked rewards $e(\mathcal M):=\mathbb E[P]$, with expectations
over the type, the selection randomisation, and the referral capacity.
Note that $e$ counts rewards reserved, not rewards eventually paid.
Since $1\le q_\ell\le\overline\tau_\ell$ for every contract, we have
$1\le u\le e$ and $0\le r\le u$.

\begin{definition}[Expected-budget relaxation]
\label{def:expected-budget-relaxation}
The relaxed recruitment of a fixed menu is
\begin{equation}
 \Phi(\mathcal M;B_1,B_2)
 :=r(\mathcal M)\min\left\{
 n,\frac{B_1}{u(\mathcal M)},\frac{B_2}{e(\mathcal M)}
 \right\},
 \label{eq:menu-relaxation}
\end{equation}
abbreviated to $\Phi(\mathcal M)$ when the budgets are clear.
\end{definition}

The minimum is the (fractional) number of individuals the budgets
could serve if each consumed exactly the mean resources. The next
theorem shows that this relaxation is accurate for the sequential
process.

\begin{theorem}[Relaxation and sequential recruitment]
\label{thm:relaxation-transfer}
Every fixed menu of admissible contracts satisfies
$\operatorname{Rec}(\mathcal M;B_1,B_2)\le\Phi(\mathcal M;B_1,B_2)$.
If every offered contract has maximum payment at most $H$, then for
$B>0$,
\begin{equation}
 \Bigl(1-\sqrt{H/B}-H/B\Bigr)\,\Phi(\mathcal M;B_1,B_2)
 \le\operatorname{Rec}(\mathcal M;B_1,B_2).
 \label{eq:relaxation-transfer}
\end{equation}
Thus, $\operatorname{Rec}=(1-o(1))\Phi$ for fixed $H$.
\end{theorem}

The two mechanism families we construct have maximum payments at most
\(m\), so their recruitment can be analysed through \(\Phi\) at a
\(1-o(1)\) loss. The proof appears in \cref{app:relaxation-transfer}.

\section{Recruitment-Maximising Mechanisms}
\label{sec:recruitment-maximizing}

With the model in place, a natural objective is to find a menu that
maximises expected recruitment for the given population and budgets.
Three questions guide our approach: how difficult such a menu is to
choose from, how difficult it is to compute, and how much recruitment
it loses to incentive constraints.

\begin{itemize}
\item \textbf{Menu complexity.}
Choosing a contract requires an individual to compare expected
payments under different reward schedules using their own referral
prospects. An optimal menu need not make these comparisons simple:
even a small instance can require several payment schedules for the
same voucher quantity.

\begin{example}[Four contracts for three outcomes]
\label{ex:four-contracts}
Consider \(m=2\) and the four types
\[
 \begin{aligned}
 D_1&=\tfrac12\delta_0+\tfrac12\delta_2,
 &D_2&=\tfrac14\delta_0+\tfrac7{32}\delta_1+\tfrac{17}{32}\delta_2,\\
 D_3&=\tfrac38\delta_1+\tfrac58\delta_2,
 &D_4&=\delta_2,
 \end{aligned}
\]
with population shares \((2/5,1/5,1/5,1/5)\). Fix
\(0<\varepsilon<1/64\), for example \(\varepsilon=1/128\), and let
\(n=5\), \(B_1=6\), and \(B_2=51/8+4\varepsilon\).
The following menu maximises relaxed recruitment:
\begin{center}
\begin{tabular}{@{}ccccc@{}}
\toprule
Contract & Vouchers \(q\) & \(\tau(0)\) & \(\tau(1)\) & \(\tau(2)\) \\
\midrule
\(C_1\) & 1 & \(1+\varepsilon\) & \(1+\varepsilon\) & \(1+\varepsilon\) \\
\(C_2\) & 1 & \(7/8-\varepsilon\) & \(9/8+\varepsilon\) & \(9/8+\varepsilon\) \\
\(C_3\) & 1 & 0 & \(5/4+\varepsilon\) & \(5/4+\varepsilon\) \\
\(C_4\) & 2 & 0 & 0 & 2 \\
\bottomrule
\end{tabular}
\end{center}
Type \(D_i\) strictly prefers \(C_i\), and relaxed recruitment is
\(19/4\). Every relaxed-optimal menu requires at least four
contracts, although there are only three possible referral outcomes;
the proof is in \cref{app:menu-example}.
\end{example}

The three one-voucher contracts in \cref{ex:four-contracts} range from
a guaranteed payment to a payment conditional on success. Thus even with only two available
quantities, an optimal menu can require participants to compare
several distinct reward schedules.

\item \textbf{Computational complexity.}
Finding an optimal menu is computationally hard, even when the
population contains only finitely many explicitly specified types
and the maximum referral capacity is a small constant.

\begin{theorem}[Hardness of recruitment maximisation]
\label{thm:menu-hardness}
Computing a recruitment-maximising admissible fixed menu is strongly
NP-hard, both for the expected-budget relaxation and for the
sequential recruitment process. This holds even when \(m=9\),
\(B_1=B_2\), and the population distribution has finite rational
support. Every contract must allocate at least one voucher, and
the result allows favourable randomisation among utility-maximising
contracts.
\end{theorem}

The proof in \cref{app:menu-hardness} reduces from
\textsc{Vertex Cover}: choosing which contracts to offer encodes
a vertex cover, while a further type enforces the necessary reward
bound.

\item \textbf{The cost of incentive constraints.}
Finally, maximising recruitment within the class of menus does not
tell us whether incentive-based allocation is the right approach
for a given population. Even the best menu may lose substantial
recruitment because participants must voluntarily choose their
assigned contracts. As we show in \cref{sec:general}, there are
populations for which every admissible fixed menu achieves only an
\(O(1/\sqrt m)\) fraction of the recruitment attainable with type
information. In such regimes, an imperfect predictor of referral
prospects, for example one learned from participant characteristics,
may support better allocations than any menu, provided its
predictions preserve enough information to guide allocation.
\end{itemize}

The first two questions motivate simple mechanisms with transparent
payment rules and efficiently computable parameters. To address the
third, we next introduce an oracle benchmark that measures the
recruitment attainable with type information and without incentive
constraints, and we compare our mechanisms with it throughout.

\section{Oracle Benchmark}
\label{sec:oracle-benchmark}

We begin with the benchmark against which all mechanisms are measured.
Consider the best expected recruitment when the designer can choose
which contract each individual receives, without enforcing incentive
constraints. The designer observes each reached individual's type
\(D\), but not their realised capacity, and uses a fixed, possibly
randomised assignment rule under the same sequential process. We
denote this optimum by \(\mathrm{OPT}(B)\). Comparing strategic
outcomes with an informed central optimum is standard in algorithmic
game theory
\citep{koutsoupias1999worst,nisan2001algorithmic,roughgarden2002selfish,procaccia2013approximate}.

Without incentive constraints, any contract assigning \(q\) vouchers
can be replaced by the linear contract \(\tau_q(z)=\min\{z,q\}\),
which preserves recruitment and reduces the earmarked reward to the
minimum admissible amount \(q\). The oracle problem is therefore
equivalent to choosing an allocation rule under a single budget of
\(B\) vouchers, which is why the two budgets enter
\(\mathrm{OPT}\) only through \(B\).

Formally, let \(\mathcal A\) be the class of fixed, possibly
randomised quantity rules \(A(D)\in\mathcal Q\), with randomisation
independent across individuals and of capacities. The oracle assigns
\(A(D)\) vouchers and stops when the remaining budget cannot cover
the largest quantity assigned by \(A\). Writing
\(\operatorname{Rec}(A;B)\) for its expected recruitment, we have
\(\mathrm{OPT}(B)=\sup_{A\in\mathcal A}\operatorname{Rec}(A;B)\).
With \(u(A):=\mathbb E[A(D)]\) and
\(r(A):=\mathbb E[v_D(A(D))]\), its relaxed value and the optimal
relaxed benchmark are
\[
 \Phi(A;B):=r(A)\min\left\{n,\frac{B}{u(A)}\right\},
 \qquad F(B):=\sup_{A\in\mathcal A}\Phi(A;B).
\]
Here \(e(A)=u(A)\), and \cref{thm:relaxation-transfer} applies
because the assignment is independent of whether an individual is
reached. Every admissible fixed menu therefore satisfies
\[
 \operatorname{Rec}(\mathcal M;B_1,B_2)
 \le\mathrm{OPT}(B)\le F(B).
\]

\subsection{Optimal Threshold Rules}
\label{sec:threshold-simple}

We now characterise the oracle's optimal allocation. The marginal recruitment gains \(p_k(D)\) are nonincreasing in
\(k\). A threshold rule therefore allocates additional vouchers as long
as these gains exceed a cutoff common to all types.

\begin{definition}[Threshold allocation rule]
\label{def:threshold-positive}
For \(\theta,\eta\in[0,1]\), the threshold rule
\(A_{\theta,\eta}\) gives every reached individual one voucher,
then includes each additional voucher \(k\in\{2,\ldots,m\}\)
with \(p_k(D)>\theta\). With probability \(\eta\), it also
includes all additional vouchers with \(p_k(D)=\theta\);
otherwise it includes none of these tied vouchers. The tie
randomisation is independent across individuals.
\end{definition}

We now show that a threshold rule is exactly optimal for the
expected-budget relaxation and hence asymptotically optimal for
sequential recruitment. \Cref{alg:oracle-threshold}
chooses its cutoff and tie probability from
\((n,B_1,B_2,\mathfrak D)\).

\begin{algorithm}[!htbp]
\caption{Choose the oracle threshold parameters}
\label{alg:oracle-threshold}
\small
\textbf{Input:} \(n,B_1,B_2,\mathfrak D\).\\
\textbf{Output:} The cutoff and tie probability
\((\theta^\star,\eta^\star)\in[0,1]^2\).
\begin{enumerate}[1.]
\setlength{\itemsep}{3pt}
\item Set \(B=\lfloor\min\{B_1,B_2\}\rfloor\). For
\(t\in[0,1]\), define the mean allocations excluding and including
threshold ties:
\[
 u^-(t)=1+\mathbb E\!\left[\sum_{k=2}^m\mathbf1\{p_k(D)>t\}\right],
 \qquad
 u^+(t)=1+\mathbb E\!\left[\sum_{k=2}^m\mathbf1\{p_k(D)\ge t\}\right].
\]
\item Find the unique zero \(\rho\in[0,1]\) of the decreasing function
\[
 G(t)=\mathbb E\!\left[p_1(D)-t+\sum_{k=2}^m(p_k(D)-t)_+\right].
\]
\item Set
\(\displaystyle u^\star=\max\{u^-(\rho),\min\{B/n,u^-(0)\}\}\).
\item Set
\(\displaystyle\theta^\star=\inf\{t\in[0,1]:u^-(t)\le u^\star\}\).
\item Set
\[
 \eta^\star=
 \begin{cases}
 \displaystyle\frac{u^\star-u^-(\theta^\star)}
 {u^+(\theta^\star)-u^-(\theta^\star)},
 &u^+(\theta^\star)>u^-(\theta^\star),\\[6pt]
 0,&u^+(\theta^\star)=u^-(\theta^\star).
 \end{cases}
\]
\item Return \((\theta^\star,\eta^\star)\).
\end{enumerate}
\end{algorithm}

\FloatBarrier
For a finite rational prior, \(G\) is piecewise affine with
breakpoints among the marginal probabilities \(p_k(D)\), \(k\ge2\).
Sorting these breakpoints together with \(0\) and \(1\), and locating
the interval containing the zero, computes \(\rho\); the same sorted list determines
\(\theta^\star\) and \(\eta^\star\). Thus
\cref{alg:oracle-threshold} runs in polynomial time in the finite
prior's input size and \(m\).

\begin{theorem}[Optimality for the expected-budget relaxation]
\label{thm:simple-threshold}
\label{lem:threshold-relaxation}
The parameters \((\theta^\star,\eta^\star)\) returned by
\cref{alg:oracle-threshold} define a rule
\(A^\star=A_{\theta^\star,\eta^\star}\) with mean
\(u(A^\star)=u^\star\) that is optimal for the expected-budget
relaxation:
\[
 \Phi(A^\star;B)=\max_{A\in\mathcal A}\Phi(A;B)=F(B).
\]
For fixed \(m\), it therefore achieves a \(1-o(1)\) approximation
to \(\mathrm{OPT}(B)\) as \(B\to\infty\).
\end{theorem}

\begin{proof}
For a type \(D\) and cutoff \(t\), the first voucher is mandatory;
each additional voucher contributes \(p_k(D)-t\) to
\(v_D(q)-tq\). Since these contributions are nonincreasing in \(k\),
a maximising prefix includes all positive terms, excludes all
negative terms, and may include terms equal to zero. Thus
\[
 \max_{q\in\mathcal Q}\{v_D(q)-tq\}
 =p_1(D)-t+\sum_{k=2}^m(p_k(D)-t)_+.
\]
The smallest and largest such choices have means \(u^-(t)\) and
\(u^+(t)\). Varying the cutoff and mixing tied choices attains every
mean \(u\in[1,m]\). If \(\widehat A_u\) is the resulting threshold
rule with cutoff \(t\), pointwise maximisation gives
\(r(\widehat A_u)-tu\ge r(A)-tu\) for any other rule with mean \(u\).
Hence \(R(u):=r(\widehat A_u)\) is the greatest recruitment at that
mean. As additional vouchers are taken in decreasing order of
marginal gain, \(R\) is continuous, nondecreasing, and concave.

Taking expectations gives
\(G(t)=\max_A\{r(A)-tu(A)\}\). Since \(u(A)\ge1\), \(G\) is
continuous and strictly decreasing, with \(G(0)\ge0\ge G(1)\).
Its zero \(\rho\) is therefore the largest recruitment per voucher,
attained at the smallest mean \(u^-(\rho)\). Concavity implies that
\(R(u)/u\) is nonincreasing thereafter. To maximise
\(R(u)\min\{n,B/u\}\), there are three cases:
\begin{enumerate}[(i)]
\setlength{\itemsep}{2pt}
\item If \(B/n\le u^-(\rho)\), the budget limits how many
individuals can be reached. Set \(u^\star=u^-(\rho)\), attaining
the maximum recruitment per voucher and relaxed value \(B\rho\).
\item If \(u^-(\rho)<B/n<u^-(0)\), set \(u^\star=B/n\).
The relaxation then serves all \(n\) individuals and uses the whole
budget: a smaller mean cannot increase \(R(u)\), while a larger
mean cannot increase \(R(u)/u\).
\item If \(B/n\ge u^-(0)\), set \(u^\star=u^-(0)\).
Every individual receives all vouchers with positive recruitment
gain, and further vouchers cannot improve recruitment.
\end{enumerate}
These cases give step 3. Steps 4--5 choose a cutoff with
\(u^-(\theta^\star)\le u^\star\le u^+(\theta^\star)\) and mix
the two endpoint choices to obtain exactly \(u^\star\). Thus
\(\Phi(A^\star;B)=F(B)\). Finally, quantities and maximum payments
are at most \(m\), so \cref{thm:relaxation-transfer} gives, for
\(B>0\), recruitment at least
\((1-\sqrt{m/B}-m/B)F(B)\). Since
\(\mathrm{OPT}(B)\le F(B)\), this proves the asymptotic claim.
\end{proof}

The threshold rule \(A^\star\) is thus a simple benchmark for
designing and analysing menus. Because it attains \(F(B)\), it
suffices to compare a menu's relaxed recruitment with that of
\(A^\star\) and then apply \cref{thm:relaxation-transfer}.

\section{Mechanism Design for General Types}
\label{sec:general}
\label{sec:warmup}
We now turn from the benchmark to mechanisms that participants choose
from voluntarily. For arbitrary referral distributions, a menu can induce the oracle's
threshold allocation by reserving additional rewards. We quantify
this cost and obtain asymptotically optimal recruitment when the
reward budget covers it, with strong guarantees when the cost is
small. We then show that, without restrictions on types, no
worst-case approximation ratio can be bounded away from zero
independently of \(m\).

\subsection{Voucher Price and Linear Payment}
\label{sec:easier-regimes}

This family of mechanisms pays linearly for successful referrals and adds a
base payment that decreases with the number of vouchers requested.

\begin{definition}[Voucher-price and linear-payment menus]
\label{def:threshold-menu-family}
For \(\theta\in[0,1]\), the mechanism \(\mathcal T_\theta\)
offers the menu
\[
 \mathcal T_\theta:=\{(q,\tau_q^\theta):q\in\mathcal Q\},
 \qquad \tau_q^\theta(z):=\min\{z,q\}+\theta(m-q).
\]
\end{definition}

\begin{figure}[!htbp]
\centering
\includegraphics[width=0.9\textwidth]{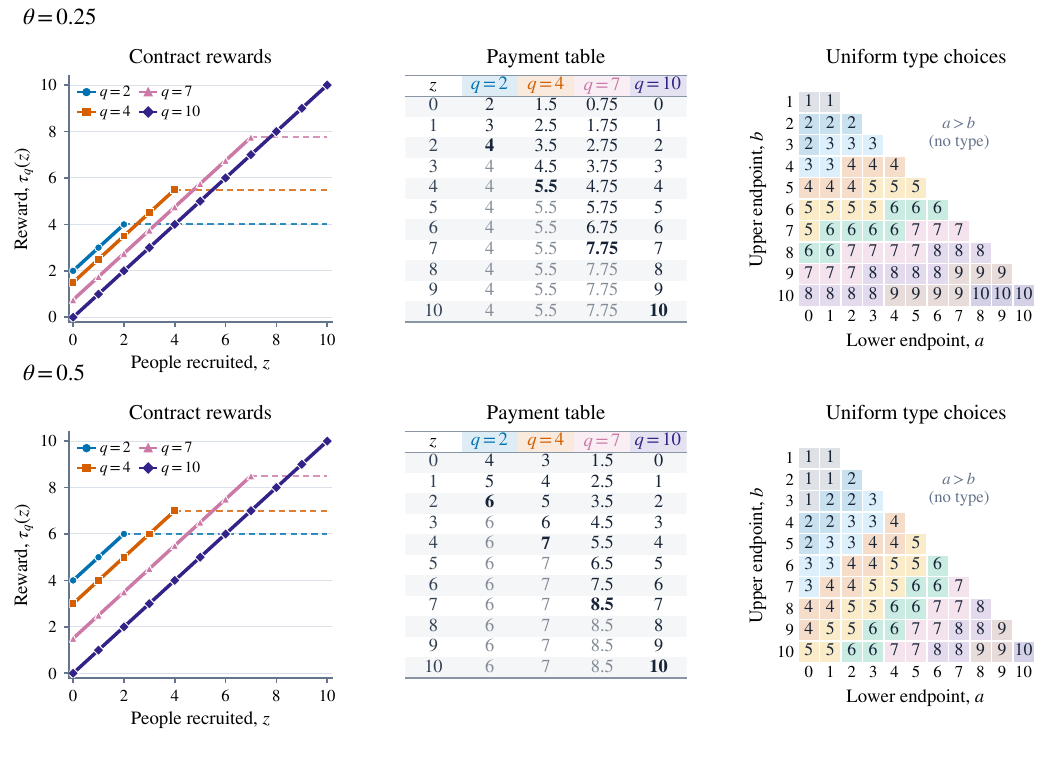}
\caption{Voucher-price and linear-payment menus with \(m=10\):
\(\theta=0.25\) (top) and \(\theta=0.5\) (bottom). In each row, the
left panel plots the rewards of the contracts \(q=2,4,7,10\); the
middle table lists them, with contracts as columns and the number
recruited \(z=0,\ldots,10\) as rows; and the right panel shows the
choices of uniform types from the full menu. Bold entries mark
completion \(z=q\); grey entries give the constant extension for
\(z>q\). Choice diagrams break
ties towards the smallest quantity.}
\label{fig:threshold-menus}
\end{figure}

The name reflects a pricing interpretation: participants start with
one voucher and a credit of \(\theta(m-1)\), spend \(\theta\) of this
credit on each additional voucher, and receive the remaining credit
plus one unit per successful referral
(\cref{fig:threshold-menus}). Each contract is admissible, with
maximum reward \(q+\theta(m-q)\le m\).

\Cref{alg:general-menu} uses the oracle's cutoff as the voucher
price and its allocation rule to resolve ties.

\begin{algorithm}[!htbp]
\caption{Construct the voucher-price menu}
\label{alg:general-menu}
\small
\textbf{Input:} \(n,B_1,B_2,\mathfrak D\).\\
\textbf{Output:} The menu \(\mathcal T_{\theta^\star}\) and its
selection rule.
\begin{enumerate}[1.]
\setlength{\itemsep}{3pt}
\item Run \cref{alg:oracle-threshold} on \((n,B_1,B_2,\mathfrak D)\)
to obtain \((\theta^\star,\eta^\star)\).
\item Construct
\[
 \mathcal T_{\theta^\star}:=
 \left\{\left(q,z\mapsto\min\{z,q\}+\theta^\star(m-q)\right):
 q\in\mathcal Q\right\}.
\]
\item Return \(\mathcal T_{\theta^\star}\) with type \(D\) selecting
quantity \(A_{\theta^\star,\eta^\star}(D)\), independently across
individuals.
\end{enumerate}
\end{algorithm}

\begin{theorem}[Threshold-dependent approximation]
\label{thm:threshold-reward-guarantee}
\label{prop:warmup-ample}
For fixed \(m\) and any population, let \(\mathcal T_{\theta^\star}\)
be the menu returned by \cref{alg:general-menu}, and put
\(\beta:=B_2/B_1\). With favourable tie-breaking, this mechanism
achieves an approximation factor of
\[
 \min\left\{1,
 \frac{\max\{1,\beta\}}{1+\theta^\star(m-1)}\right\}-o(1)
\]
to \(\mathrm{OPT}(B)\) as \(B\to\infty\).
\end{theorem}

\begin{proof}
Let \(\mathcal M=\mathcal T_{\theta^\star}\). Type \(D\)'s
utility from quantity \(q\) is
\(U_q(D)=v_D(q)+\theta^\star(m-q)\), so
\[
 U_k(D)-U_{k-1}(D)=p_k(D)-\theta^\star
 \qquad (2\le k\le m).
\]
Since the tails are nonincreasing, a best response includes every
additional voucher \(k\ge2\) with \(p_k(D)>\theta^\star\) and none with
\(p_k(D)<\theta^\star\). Every quantity assigned by \(A^\star\) is
therefore a best response. Favourable tie-breaking selects exactly
as \(A^\star\) does, including its randomisation at the threshold.
Write \(r^\star:=r(A^\star)\) and \(u^\star:=u(A^\star)\) for
the resulting mean recruitment and voucher use. The mean earmarked
reward is
\[
 e^\star=u^\star+\theta^\star(m-u^\star)=w^\star u^\star,
 \qquad w^\star:=1+\theta^\star(m/u^\star-1)
 \le1+\theta^\star(m-1).
\]
Put \(\gamma=\min\{1,\max\{1,\beta\}/w^\star\}\).
Since \(B_1\ge B\) and \(B_2\ge B\max\{1,\beta\}\),
\begin{equation}
 \Phi(\mathcal M)
 =r^\star\min\left\{n,\frac{B_1}{u^\star},
                         \frac{B_2}{w^\star u^\star}\right\}
 \ge\gamma r^\star\min\{n,B/u^\star\}=\gamma F(B).
 \label{eq:mean-aware-interpolation}
\end{equation}
Maximum payments are at most \(m\), so
\cref{thm:relaxation-transfer} and
\(F(B)\ge\mathrm{OPT}(B)\) finish the proof.
\end{proof}

\begin{remark}[When the mechanism is asymptotically optimal]
\label{rem:easy-regimes}
The guarantee is \(1-o(1)\) whenever
\(B_2\ge[1+\theta^\star(m-1)]B_1\); in particular,
\(B_2\ge mB_1\) always suffices.
Furthermore, if \(\theta^\star(m-1)\le\varepsilon\),
the factor is at least \(1/(1+\varepsilon)-o(1)\).
\end{remark}

\begin{remark}[The price of information]
\label{rem:price-of-information}
Compared with the oracle, which reserves \(q\), the menu reserves an
extra \(\theta^\star(m-q)\) to elicit the threshold choice. The
largest such amount, \(\theta^\star(m-1)\), can be viewed as the
\emph{price of information} of this mechanism. When \(B_2>B_1\),
the surplus reward budget pays part of this price, and the factor
reaches one once the surplus covers all of it. A small threshold
\(\theta^\star\) makes the same information cheaper to elicit.
\end{remark}

\subsection{Impossibility for Unrestricted Types}

The guarantee of \cref{thm:threshold-reward-guarantee} degrades as the
price of information \(\theta^\star(m-1)\) grows relative to
\(B_2/B_1\). We next show that this loss is not specific to the
voucher-price mechanism: without a reward surplus, the cost of incentives can rule out any
approximation ratio independent of \(m\).

\begin{proposition}[Impossibility for unrestricted types]
\label{prop:warmup-impossibility}
\label{prop:unrestricted-sharper}
For unrestricted types, the best worst-case approximation ratio is
\(O(1/\sqrt m)\) in the large-budget limit. For every \(m\ge2\)
and integer \(B>m\), there is an instance with
\(n=B_1=B_2=B\) on which every admissible fixed menu satisfies
\[
 \frac{\operatorname{Rec}(\mathcal M;B,B)}{\mathrm{OPT}(B)}
 \le\frac{5B}{2\sqrt m\,(B-m)}.
\]
\end{proposition}

\begin{proof}[Proof sketch]
For sufficiently large \(m\), take \(n=B_1=B_2=B\) and the
following population; the final column gives the oracle rule used
for comparison.
\begin{center}
\small
\begin{tabular}{@{}lccc@{}}
\toprule
Type & Capacity distribution & Population share & Oracle vouchers \\
\midrule
Reliable \(H\) & \(\delta_m\) & \(m^{-1}\) & \(m\) \\
Occasional \(L\) & \((1-m^{-1/2})\delta_0+m^{-1/2}\delta_m\)
  & \(m^{-1/4}\) & \(1\) \\
Zero capacity & \(\delta_0\) & \(1-m^{-1}-m^{-1/4}\) & \(1\) \\
\bottomrule
\end{tabular}
\end{center}
The oracle concentrates vouchers on the rare reliable recruiters.
A menu must instead induce them to request these vouchers, yet a
large completion reward also attracts occasional recruiters.
Giving the latter a preferred alternative requires either reserving
large rewards for them or offering base payments that also appeal
to zero-capacity participants. These incentive costs limit the
menu's recruitment to an \(O(m^{-1/2})\) fraction of the oracle's.
The full proof appears in \cref{app:unrestricted-proof}.
\end{proof}

\section{Mechanism Design for Log-Concave Types}
\label{sec:restricted}
\label{sec:uniform-mechanisms}
\label{sec:linear-mechanisms}

The construction behind \cref{prop:warmup-impossibility} relies on a
polarised referral distribution: an occasional recruiter recruits either
nobody or the full capacity \(m\). We therefore ask whether constant
approximation guarantees hold for more regular referral distributions,
and study the following family.

\begin{definition}[Log-concave types]
\label{def:lc-types}
A type \(D\) is \emph{log-concave} if its support is an integer
interval and \(D(j)^2\ge D(j-1)D(j+1)\) for all \(1\le j\le m-1\).
\end{definition}

Log-concave types include the following probability distributions,
which capture several plausible patterns of recruitment:
\begin{itemize}
\setlength{\itemsep}{1pt}
\item \emph{Uniform:} the participant knows roughly how many people
they can recruit (for example, ``I can recruit about \(3\) to \(5\)
people'') but cannot say which number in that range is more likely.
\item \emph{Binomial:} the participant will approach \(N\le m\) contacts,
each of whom takes part independently with probability \(p\).
\item \emph{Truncated geometric:} the participant keeps recruiting
until the first refusal, with each attempt succeeding independently
with the same probability, conditional on recruiting at most \(m\)
people.

\end{itemize}

The polarised types behind \cref{prop:warmup-impossibility} are not
log-concave, since their support skips the capacities strictly between
\(0\) and \(m\). We use log-concavity as one natural way to formalise
regular referral distributions. It restricts each participant's own
distribution, not the population: the prior over log-concave types is
arbitrary, so participants may still differ widely in how many
contacts they can recruit. We introduce a one-parameter family of menus
(\cref{sec:menu-constructions}), prove a constant-factor guarantee for
log-concave types (\cref{sec:shared-proof}), and show that this
guarantee is tight (\cref{sec:restricted-upper}).

\subsection{Linear Payment and Completion Bonus}
\label{sec:menu-constructions}

The voucher-price menu reproduces the oracle's allocation by reserving
additional rewards. We now construct menus that reserve exactly one
reward unit per voucher, as the oracle does, and approximate the
oracle through the allocation they induce. The two simple payout rules
of \cref{ex:intro-linear,ex:intro-all-or-nothing} err in opposite
directions: linear payments encourage excessive requests, whereas
all-or-nothing payments discourage useful ones. Our menus interpolate
between them.

\begin{definition}[Linear-payment and completion-bonus menus]
\label{def:completion-menu-family}
For \(\lambda\in[0,1]\), the mechanism \(\mathcal C_\lambda\)
offers the menu \(\mathcal C_\lambda:=
\{(q,\tau_q^\lambda):q\in\mathcal Q\}\), where
\begin{equation}
 \tau_q^\lambda(z)
 :=(1-\lambda)q\mathbf 1\{z\ge q\}
                  +\lambda\min\{z,q\}.
\label{eq:completion-linear-menu}
\end{equation}
\end{definition}

\begin{figure}[!htbp]
\centering
\includegraphics[width=0.9\textwidth]{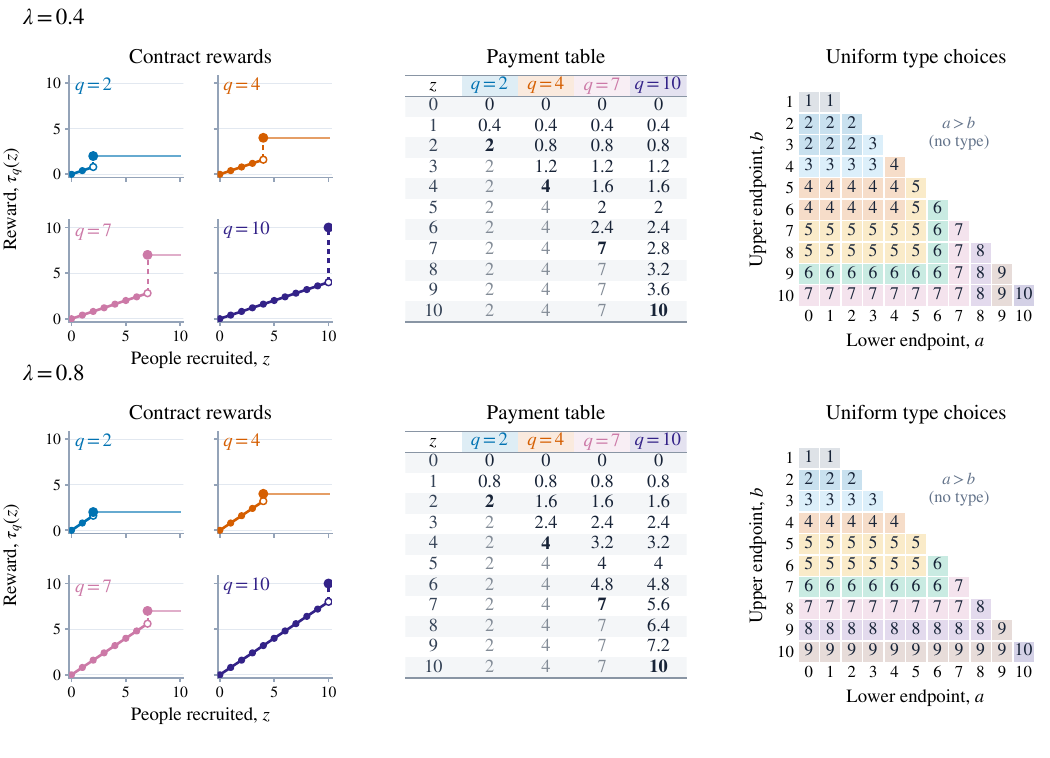}
\caption{Linear-payment and completion-bonus menus with \(m=10\):
\(\lambda=0.4\) (top) and \(\lambda=0.8\) (bottom). Panels are as in
\cref{fig:threshold-menus}: contract rewards and the payment table for
\(q=2,4,7,10\), and the choices of uniform types from the full menu of
ten contracts. Dashed vertical segments mark completion bonuses. Bold
table entries mark \(z=q\); grey entries give the constant extension
for \(z>q\). Choice diagrams 
break ties towards the smallest quantity.}
\label{fig:completion-mixtures}
\end{figure}

At \(\lambda=0\), this is the \emph{all-or-nothing rule}:
pay \(q\) only upon completing all \(q\) referrals.
At \(\lambda=1\), it is the \emph{linear rule}: pay one unit per
successful referral, up to \(q\). Intermediate weights combine a
payment per referral with a completion bonus within each contract.
Every contract earmarks exactly \(q\), the same amount as the oracle,
and has nondecreasing payments, so every menu in the family is
admissible.

Under \(\mathcal C_\lambda\), type \(D\)'s expected payment from
quantity \(q\) is
\(
 (1-\lambda)\,q\,p_q(D)+\lambda\,v_D(q),
\)
the sum of the expected completion bonus and the expected linear
payment. At \(\lambda=0\), type \(D\) requests
\[
 k_0(D):=\min\argmax_{q\in\mathcal Q}\{q\,p_q(D)\},
 \qquad\text{with population mean } u_0:=\mathbb E[k_0(D)];
\]
for example, a uniform type on \(\{a,\ldots,b\}\) requests
\(k_0(D)=\max\{1,a,\lfloor(b+1)/2\rfloor\}\). Raising \(\lambda\)
shifts weight towards the linear payment, which rewards every used
voucher, and therefore encourages larger requests, up to the largest
useful quantity \(\bar b(D)\). The next proposition makes this
precise for best responses between \(k_0(D)\) and \(\bar b(D)\),
which are the ones we select throughout this construction. It holds
for every population distribution; its formal statement and proof
appear in \cref{app:menu-attainability}.

\begin{proposition}[Monotonicity and attainable mean allocations]
\label{lem:shared-attainability}
As \(\lambda\) increases from \(0\) to \(1\), each type's best
responses under \(\mathcal C_\lambda\) are nondecreasing, from its
all-or-nothing choice \(k_0(D)\) up to its largest useful quantity
\(\bar b(D)\). Consequently, with favourable tie-breaking, the menus
\(\mathcal C_\lambda\) attain every mean allocation
\(u\in[u_0,u_{\max}]\), where \(u_{\max}:=\mathbb E[\bar b(D)]\): for
each such \(u\), there is a smallest weight \(\lambda_u\) at which some
selection of best responses has mean at least \(u\), and
\(\mathcal C_{\lambda_u}\) admits a selection with mean exactly \(u\).
\end{proposition}

\Cref{alg:restricted-menu} uses this to match the oracle's mean
allocation \(u^\star\) whenever all-or-nothing payments fall short of it.

\begin{algorithm}[!htbp]
\caption{Construct a linear-payment and completion-bonus menu}
\label{alg:restricted-menu}
\small
\textbf{Input:} \(n,B_1,B_2,\mathfrak D\).\\
\textbf{Output:} A menu \(\mathcal M^\star\) and its selection rule.
\begin{enumerate}[1.]
\setlength{\itemsep}{3pt}
\item Compute \(u^\star\) using steps 1--3 of
\cref{alg:oracle-threshold}, and compute \(u_0=\mathbb E[k_0(D)]\).
\item \textbf{If} \(u_0\ge u^\star\), return the all-or-nothing menu
\(\mathcal C_0\) with type \(D\) selecting \(k_0(D)\).
\item Otherwise, set \(\lambda^\star:=\lambda_{u^\star}\), the smallest
weight given by \cref{lem:shared-attainability} at which the mean
allocation can reach \(u^\star\).
\item Return \(\mathcal C_{\lambda^\star}\) with a selection of best
responses whose mean allocation is exactly \(u^\star\)
(\cref{lem:shared-attainability}).
\end{enumerate}
\end{algorithm}

Since \(u^\star\le u_{\max}\), the weight \(\lambda^\star\) always
exists, and \cref{alg:restricted-menu} is well defined for every
population distribution, not only for log-concave types. For a
population with finite support, \(\lambda^\star\) and its selection
rule can be computed in polynomial time by scanning \(0\), \(1\), and
the intersections of each type's utility lines
(\cref{app:menu-attainability}).

\subsection{Approximation Guarantees}
\label{sec:shared-proof}

We now show that the menu returned by \cref{alg:restricted-menu}
recruits a constant fraction of the oracle's recruitment for
log-concave types.

\begin{theorem}[Approximation for log-concave types]
\label{thm:restricted-guarantees}
For fixed \(m\) and every population of log-concave types, with
favourable tie-breaking, the menu \(\mathcal M^\star\) returned by
\cref{alg:restricted-menu} achieves
\[
 \operatorname{Rec}(\mathcal M^\star;B_1,B_2)
 \ge\bigl(1-1/e-o(1)\bigr)\mathrm{OPT}(B).
\]
\end{theorem}

\paragraph{Proof idea.}
Every contract in \(\mathcal C_\lambda\) earmarks exactly its
quantity, as under the oracle, so the two budgets act as a single
budget \(B\). The key step compares two quantities for a single type.
Let \(k\) be a quantity that type \(D\) selects under
\(\mathcal C_\lambda\), and let \(q\) be any other quantity, such as
the oracle's \(A^\star(D)\). Using the shape of log-concave tails at a
best response, we show (\cref{lem:family-matching}) that
\begin{equation}
 v_D(k)-\tfrac7{10}\,v_D(q)\ \ge\ c_\lambda\,(k-q),
 \qquad c_\lambda=\tfrac{3(1-\lambda)}{10}.
 \label{eq:proof-idea-matching}
\end{equation}
Thus a type's own choice recovers a \(7/10\) fraction of the
recruitment from any other quantity, up to a correction proportional
to the quantity difference: if the type takes fewer vouchers than
\(q\), its recruitment falls short of \(\frac7{10}v_D(q)\) by at most
\(c_\lambda\) per missing voucher, and if it takes more, it exceeds
\(\frac7{10}v_D(q)\) by at least \(c_\lambda\) per extra voucher.
Setting \(q=A^\star(D)\) and averaging over types, the correction
becomes \(c_\lambda(\mathbb E[k]-u^\star)\), which disappears when the
menu's mean allocation matches the oracle's. This is why
\cref{alg:restricted-menu} calibrates \(\lambda^\star\) so that the
mean allocation equals \(u^\star\); averaging
\eqref{eq:proof-idea-matching} then gives mean recruitment at least
\(7/10\) times the oracle's.

The all-or-nothing branch needs a separate argument, and it determines
the constant. When \(u_0>u^\star\), no weight in \([0,1]\) lowers the
mean allocation to \(u^\star\), so the menu gives out more vouchers per
individual than the oracle and the budget may reach fewer individuals.
Comparing recruitment per individual is then not enough. We instead
compare recruitment per voucher, showing that all-or-nothing payments
attain at least \(1-1/e\) times the oracle's recruitment per voucher
(\cref{lem:starting-density}). Since \(1-1/e<7/10\), this branch sets
the guarantee. The proof appears in \cref{app:restricted-proof}.

\begin{remark}[Combining the guarantees]
\label{rem:combined-uniform-guarantee}
Before types are drawn, the designer can also evaluate the
voucher-price mechanism of \cref{thm:threshold-reward-guarantee} and
use whichever menu has the larger relaxed recruitment under the prior.
For log-concave types, this achieves at least
\[
 \max\left\{1-\frac1e,\
 \min\left\{1,\frac{\max\{1,\beta\}}{1+\theta^\star(m-1)}\right\}
 \right\}-o(1).
\]
\end{remark}
\FloatBarrier
\subsection{Impossibility Bounds and Tightness}
\label{sec:restricted-upper}

We conclude by showing that no admissible menu can do better on
log-concave types.

\begin{theorem}[Impossibility for log-concave types]
\label{thm:uniform-upper}
\label{thm:linear-upper}
For every \(\eta>0\), there exist \(m\), a population of
log-concave types, and a constant \(c\ge1\) such that, as
\(n\to\infty\) with \(B_1=B_2=B=\lfloor cn\rfloor\), every
admissible fixed menu satisfies
\[
\operatorname{Rec}(\mathcal M;B,B)
\le\bigl(1-1/e+\eta+o(1)\bigr)\mathrm{OPT}(B).
\]\end{theorem}

The proof, in \cref{app:upper-bounds}, uses truncated geometric types,
with \(m\) growing as \(\eta\to0\). The \(1-1/e\) guarantee of
\cref{thm:restricted-guarantees}, which holds for every \(m\),
therefore cannot be improved.

\section{Experiments}
\label{sec:experiments}

We now evaluate both mechanism families on a population prior built
from HIV transmission network data. We study how truncating \(m\) in
the voucher-price menu and the budget ratio \(B_2/B_1\) affect
recruitment (\cref{sec:exp-caps}), how far the algorithms' parameters
are from the best parameters found by tuning (\cref{sec:exp-tuning}),
how the mechanisms respond to more extreme capacity distributions
(\cref{sec:exp-extreme}), and how they respond to a misspecified prior
(\cref{sec:exp-misspecified}).

\subsection{Setup}
\label{sec:exp-setup}

\paragraph{Population prior.}
The prior is based on the HIV Transmission Network Metastudy dataset
from ICPSR \citep{morris2011hiv}, with a construction similar to that
of \citet{pan2026adaptive}. It has \(64\) types, with capacities
between \(0\) and \(m=20\); \cref{app:experiments-data} gives the
details. We do not require these empirical types to be log-concave:
log-concavity is a sufficient condition for the guarantee of
\cref{thm:restricted-guarantees}, not a property we impose on the
data. Indeed, only \(6\) of the \(64\) types, with \(8.4\%\) of
the population, are log-concave.

\paragraph{Mechanisms.}
We compare the following policies:
\begin{itemize}
\setlength{\itemsep}{2pt}
\item \textbf{Oracle threshold policy:} the rule \(A^\star\) of
\cref{alg:oracle-threshold}, which assigns vouchers directly by type.
\item \textbf{Constant allocation:} every individual receives
\(q=B_1/n\) vouchers, an integer in every setting we consider, and
is paid one unit per successful referral.
This single-contract menu uses no information about types and does
not depend on the prior, like the fixed coupon allocations common in
practice (\cref{sec:introduction}).
\item \textbf{Completion bonus:} the menu of \cref{alg:restricted-menu}.
\item \textbf{Voucher price:} the menu \(\mathcal T_{\theta^\star}\) of
\cref{alg:general-menu}. Since the price of information
\(\theta^\star(m-1)\) grows with \(m\)
(\cref{rem:price-of-information}), we also offer this menu with \(m\)
truncated to a smaller value, which we call the \emph{voucher cap}:
quantities \(q\in\{1,\ldots,m\}\) and payments
\(\min\{z,q\}+\theta^\star(m-q)\). We keep the cutoff \(\theta^\star\)
and tie probability \(\eta^\star\) computed for \(m=20\). Since the
marginal utility of voucher \(k\) is still \(p_k(D)-\theta^\star\),
each individual selects the truncated oracle choice
\(\min\{A^\star(D),m\}\), which is a best response. When a single cap
is needed, we use the \emph{best fitted cap}: the cap whose menu has
the largest relaxed recruitment \eqref{eq:menu-relaxation} under the
prior used to fit the mechanisms.
\end{itemize}

\paragraph{Recruitment and normalisation.}
Each setting has \(n=1{,}000\) individuals. The main text uses
\(B_1=5n\) and \(B_2\in\{B_1,1.5B_1\}\);
\cref{app:experiments-additional} repeats every experiment with
\(B_1\in\{4n,6n\}\). We simulate the sequential process of
\cref{sec:recruitment-process}: types are drawn independently from the
prior, each reached individual selects a contract, and recruitment
stops once a remaining budget cannot cover the whole menu. We report
mean recruitment over \(100\) trials divided by the relaxed benchmark
\(F(B)\) of the true population. Since \(F(B)\ge\mathrm{OPT}(B)\),
the reported ratios are lower bounds on the ratios to
\(\mathrm{OPT}(B)\). Error bars are pointwise \(95\%\) confidence
intervals for the mean. With a correct prior, the oracle threshold
policy attains about \(0.98\) to \(0.99\) in every setting; the
remaining gap is the finite-budget loss of
\cref{thm:relaxation-transfer}.

\subsection{Voucher Caps and the Budget Ratio}
\label{sec:exp-caps}

\begin{figure}[!htbp]
\centering
\includegraphics[width=\textwidth]{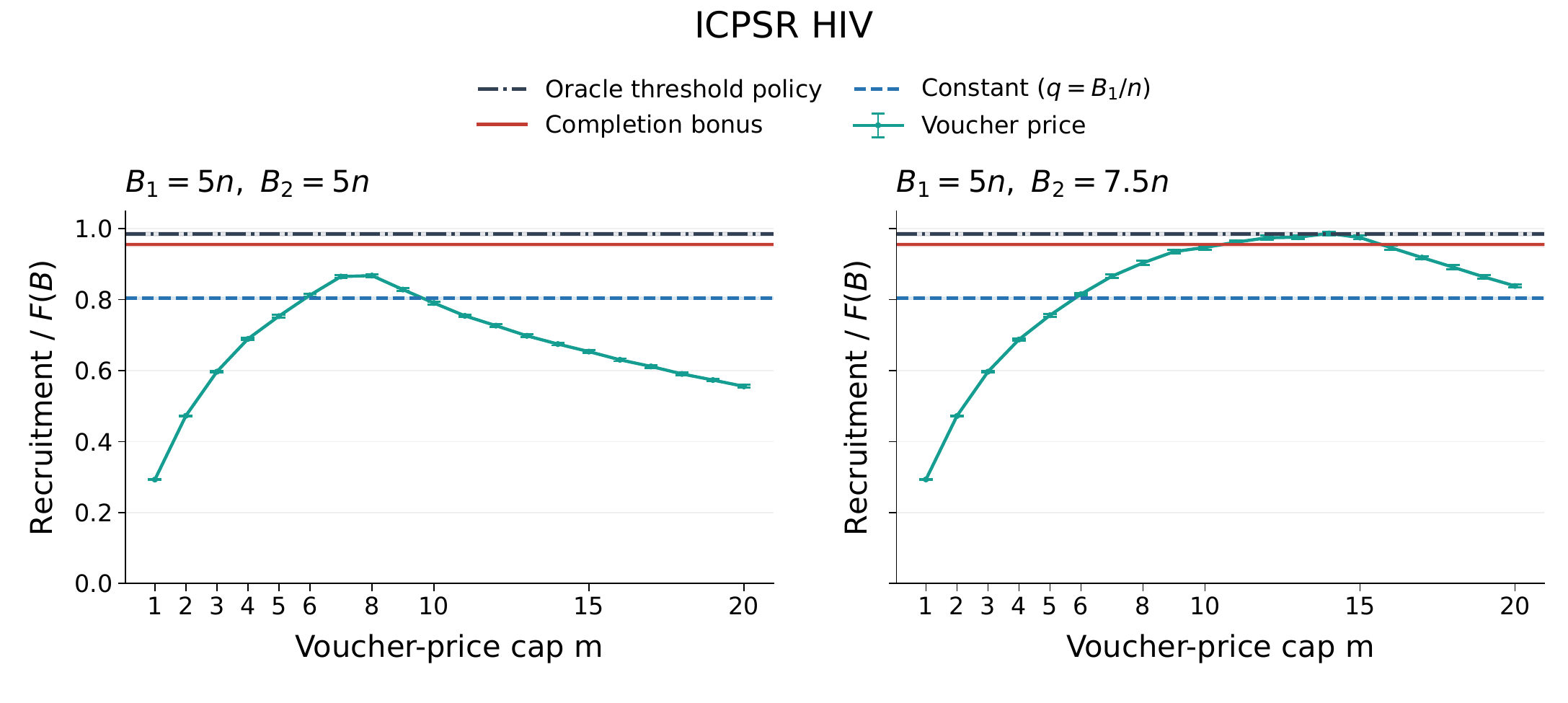}
\caption{Effect of the voucher cap, with \(n=1{,}000\), \(B_1=5n\),
and \(B_2=B_1\) (left) or \(B_2=1.5B_1\) (right). Every cap keeps the
cutoff and tie rule computed for \(m=20\); the oracle threshold
policy, the completion bonus, and the constant allocation appear as
horizontal references. Recruitment is divided by \(F(B)\) of the true
population. Error bars are pointwise \(95\%\) confidence intervals
over \(100\) trials.}
\label{fig:exp-caps}
\end{figure}

\paragraph{Capping can increase recruitment.}
In both panels of \cref{fig:exp-caps}, recruitment under the voucher
price first rises and then falls as the cap grows. At
\(B_2=B_1\), the cap \(m=8\) raises recruitment from \(0.56\) at
\(m=20\) to \(0.87\). Unlike the oracle, a capped menu cannot tap into
the recruitment potential of high-capacity individuals. However,
capping effectively decreases \(m\) and with it the price of
information \(\theta^\star(m-1)\), leaving more of the reward budget
for reaching individuals.

\paragraph{Comparison of the policies.}
The completion bonus reserves exactly \(q\) rewards for \(q\)
vouchers, as the oracle does, so it cannot use a reward surplus; its
recruitment (\(0.95\)) is the same in both panels. The voucher price
uses the surplus to pay the price of information: when
\(B_2=1.5B_1\), the cap \(m=14\) attains \(0.99\), matching the oracle
and better than the completion bonus. When \(B_2=B_1\), the completion
bonus outperforms the voucher price for every cap. Both mechanisms
outperform the constant allocation (\(0.80\)), which ignores
participants' private information: the completion bonus by about
\(15\) points of \(F(B)\), and the voucher price, with its best cap,
by about \(6\) points when \(B_2=B_1\) and about \(18\) points when
\(B_2=1.5B_1\).

\subsection{Distance from Optimal Parameters}
\label{sec:exp-tuning}

\begin{figure}[!htbp]
\centering
\includegraphics[width=\textwidth]{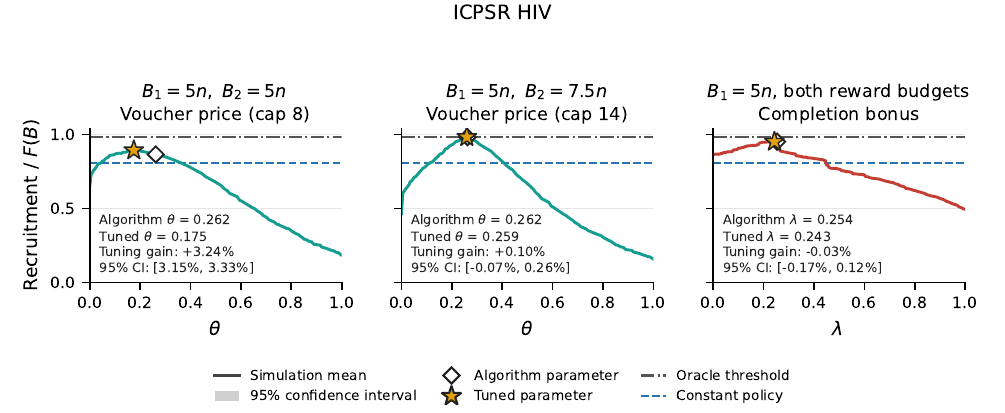}
\caption{Recruitment as a function of the voucher price \(\theta\) at
the best fitted cap, with \(B_1=5n\) and \(B_2=B_1\) (left) or
\(B_2=1.5B_1\) (middle), and of the completion-bonus weight
\(\lambda\) (right), which is the same for both reward budgets since
every contract reserves exactly its quantity. Diamonds mark the
parameters chosen by \cref{alg:general-menu,alg:restricted-menu};
stars mark the tuned parameters. Each panel reports the relative gain
from tuning with a \(95\%\) confidence interval.}
\label{fig:exp-tuning}
\end{figure}

Both algorithms choose their parameters from the prior: the voucher
price uses the oracle's cutoff \(\theta^\star\), and the completion
bonus uses the weight \(\lambda^\star\) that matches the oracle's mean
allocation. To see how far these choices are from the best
parameters our search finds, we fix the best fitted cap and tune
\(\theta\), and separately tune \(\lambda\), selecting the value with
the highest mean recruitment over \(100\) tuning trials and evaluating
it on \(100\) independent trials. The comparison is therefore with the
best values found by this search, conditional on the selected cap for
the voucher price; it does not certify a global optimum over either
family.

As \cref{fig:exp-tuning} shows, the weight \(\lambda^\star\) of the
completion bonus performs as well as the best \(\lambda\) found by tuning: tuning
changes its recruitment by \(-0.03\%\) (\(95\%\) CI
\([-0.17\%,0.12\%]\)). At the selected cap, the cutoff
\(\theta^\star\) of the voucher price likewise performs as well as the best \(\theta\) found
when \(B_2=1.5B_1\) (\(+0.10\%\), CI \([-0.07\%,0.26\%]\)). When \(B_2=B_1\), however,
lowering \(\theta\) from \(0.262\) to \(0.175\) raises recruitment by
\(3.2\%\). The cutoff \(\theta^\star\) is computed for the oracle,
which reserves no additional rewards; when the reward budget binds, a
lower price reduces the price of information at the cost of slightly
larger requests. Across all six budget settings, tuning improves the
voucher price by at most \(6.5\%\) and the completion bonus by at most
\(0.2\%\) (\cref{app:experiments-additional}). Recruitment under the completion
bonus does, however, drop by about \(6\) points of \(F(B)\) once
\(\lambda\) exceeds \(\lambda^\star\) by about \(0.01\), as many types
switch to larger quantities at once. This jump is an effect of the
prior's finite support, in which each of the \(64\) types carries
positive mass (\cref{sec:mechanisms}); the sensitivity reappears under
misspecification.

\subsection{More Extreme Types}
\label{sec:exp-extreme}

The impossibility result of \cref{prop:warmup-impossibility} relies
on types that recruit either nobody or the maximum. To mimic it, we
make every type more extreme while preserving its mean. Let \(w_j\)
and \(\mu_j\) be the population share and mean capacity of type
\(D_j\). For \(\omega\in[0,0.5]\), we replace each type by
\begin{equation}
 D_{j,\omega}
 =(1-\omega)D_j
 +\omega\Bigl[\Bigl(1-\frac{\mu_j}{20}\Bigr)\delta_0
              +\frac{\mu_j}{20}\,\delta_{20}\Bigr],
 \label{eq:exp-extreme}
\end{equation}
mixing in a component that recruits either nobody or \(20\) people
and has the same mean. Shares and means are unchanged. We refit all
mechanisms, reselect the best fitted cap, and recompute \(F(B)\) for
each \(\omega\).

\begin{figure}[!htbp]
\centering
\includegraphics[width=\textwidth]{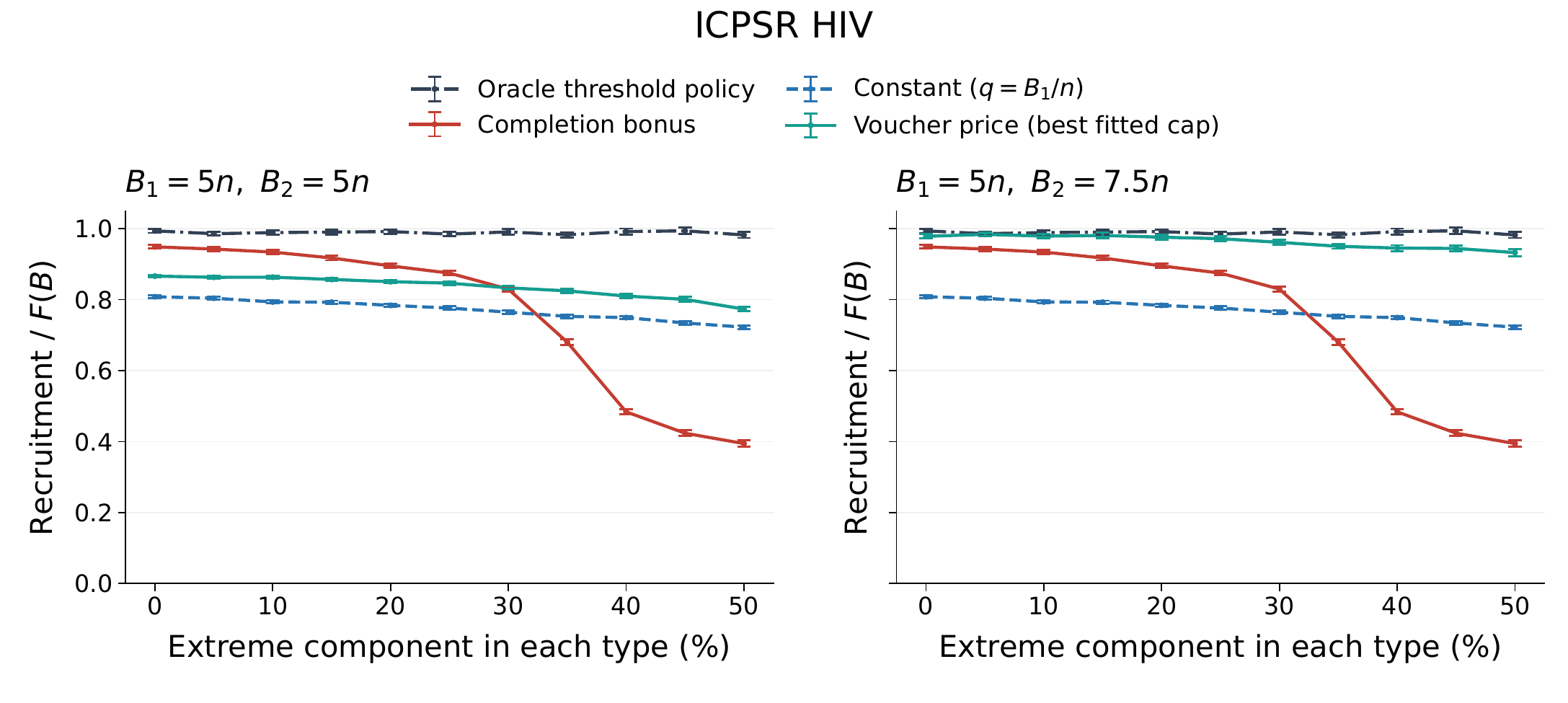}
\caption{More extreme types. Each type \(D_j\) is replaced by
\(D_{j,\omega}\) in \eqref{eq:exp-extreme}, which mixes in, with
weight \(\omega\), a component on \(\{0,20\}\) with the same mean.
All mechanisms and the best fitted cap are refitted for each
\(\omega\). Recruitment is divided by \(F(B)\) of the modified
population. Budgets and error bars are as in \cref{fig:exp-caps}.}
\label{fig:exp-extreme}
\end{figure}

As \cref{fig:exp-extreme} shows, the completion bonus falls from
\(0.95\) at \(\omega=0\) to \(0.39\) at \(\omega=0.5\), and from
\(\omega=0.35\) onwards it falls below even the constant allocation.
The voucher price declines gradually, by at most about \(0.09\),
similar to the constant allocation; for \(\omega\ge0.35\), it
outperforms the completion bonus even when \(B_2=B_1\). This reflects
a difference in what the mechanisms rely on: the completion bonus
relies on types being regular, with log-concavity as the sufficient
condition of \cref{sec:restricted}, and degrades as types
become more polarised, whereas the voucher price induces the oracle's
threshold allocation whatever the shape of the types.

\subsection{Misspecified Priors}
\label{sec:exp-misspecified}

Both mechanisms are computed from the prior, which in practice is
estimated. We misspecify the type shares by tilting them towards
types with high or low mean capacity.
Let \(s_j=(\mu_j-\min_k\mu_k)/(\max_k\mu_k-\min_k\mu_k)\in[0,1]\) be
the normalised mean of type \(j\). For a tilt strength
\(\varepsilon\), the designer uses the fitted shares
\begin{equation}
 \widehat w_j
 =\frac{w_j\exp(\varepsilon Z s_j)}{\sum_kw_k\exp(\varepsilon Z s_k)},
 \qquad Z\sim\operatorname{Unif}[0.5,1.5],
 \label{eq:exp-tilt}
\end{equation}
with \(Z\) drawn once per trial. Positive \(\varepsilon\)
overweights high-mean types, so the designer overestimates
recruitment; negative \(\varepsilon\) underestimates it. We fit every
policy, including the oracle threshold policy and the best fitted
cap, on the fitted prior and evaluate it on the true population. To
isolate the role of the cap, we also run the voucher price with a cap
three smaller than the best fitted cap (but at least one), keeping the
same fitted cutoff. The constant allocation does not depend on the
prior. The denominator, the true \(F(B)\), is the same for every
\(\varepsilon\), so changes in the plotted ratios are changes in
absolute recruitment.

\begin{figure}[!htbp]
\centering
\includegraphics[width=\textwidth]{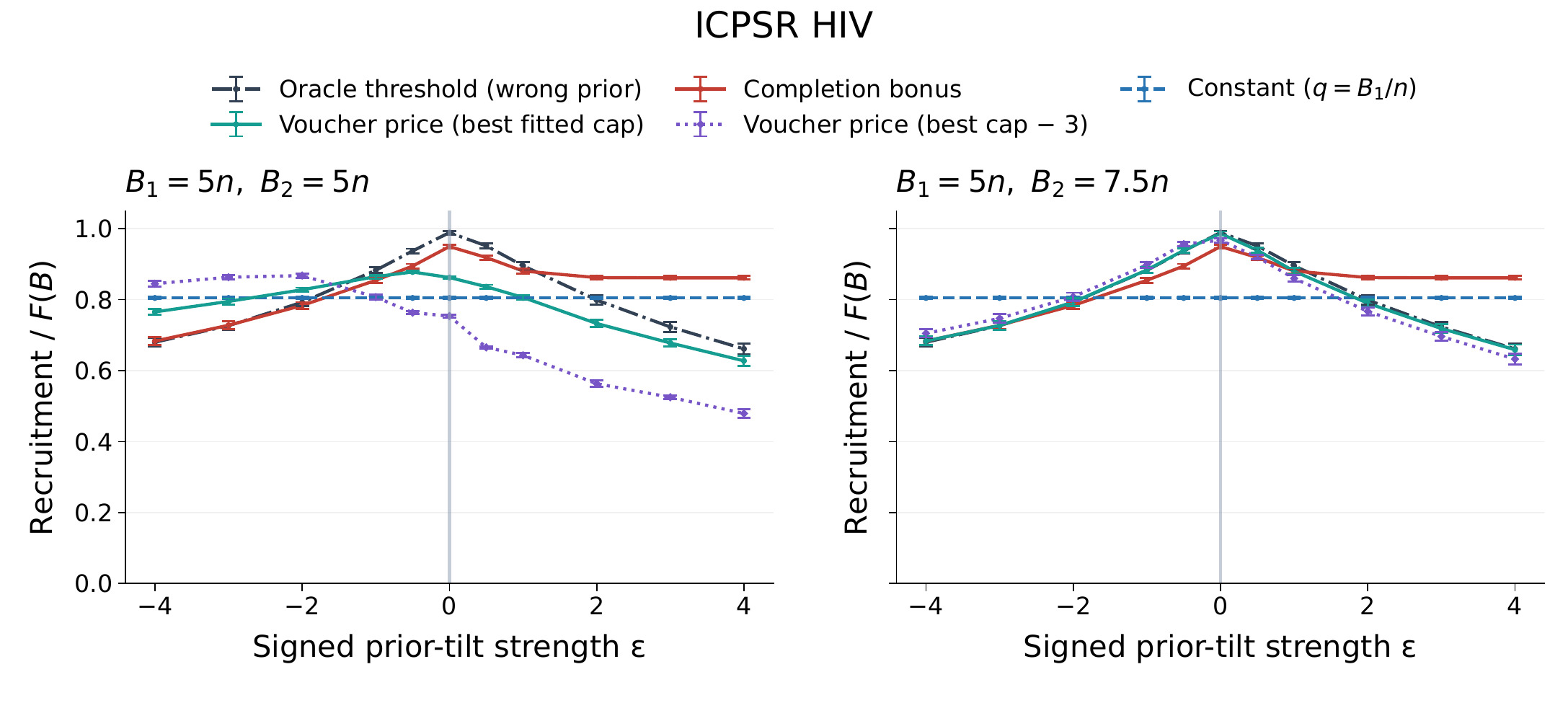}
\caption{Misspecified type shares. Every policy except the constant
allocation, including the displayed oracle threshold policy and the
best fitted cap, is fitted on the shares \eqref{eq:exp-tilt} and
evaluated on the true population. The dotted curve reduces the best
fitted cap by three, keeping the same fitted cutoff. Negative
\(\varepsilon\) underweights high-mean types; positive
\(\varepsilon\) overweights them. The denominator is the true
\(F(B)\), fixed across \(\varepsilon\). Budgets are as in
\cref{fig:exp-caps}; error bars include both prior-error and
recruitment variation.}
\label{fig:exp-prior}
\end{figure}

As \cref{fig:exp-prior} shows, the oracle threshold policy fitted on
the wrong prior degrades in both directions and falls below the
prior-free constant allocation once \(|\varepsilon|\ge2\). Each
mechanism can outperform it on one side. For \(\varepsilon\ge2\), the
completion bonus does so (\(0.86\) versus \(0.66\) at
\(\varepsilon=4\)). For \(\varepsilon\le-2\) and \(B_2=B_1\), the
voucher price with the best fitted cap does so (\(0.77\) versus
\(0.68\) at \(\varepsilon=-4\)), and reducing the cap by three raises
its recruitment further, to \(0.85\), above every other policy,
including the constant allocation. The smaller cap costs
recruitment when the prior is correct or overestimates recruitment
(\(0.75\) versus \(0.86\) at \(\varepsilon=0\)).

\paragraph{Underestimation (\(\varepsilon<0\)).}
The fitted population has too many low-mean types, so to allocate
\(u^\star\) vouchers on average to it, \cref{alg:oracle-threshold}
sets the threshold \(\widehat\theta\) too low. The oracle threshold
policy optimised for this fitted prior then gives out vouchers whose
probability of use is only slightly above \(\widehat\theta\), and on
the true population the voucher budget runs out before all
individuals are reached. Capping helps the voucher price in two ways.
First, it reduces the price of information \(\widehat\theta(m-1)\),
which the lower \(\widehat\theta\) reduces further, so the reward
budget reaches more individuals. Second, it limits these excessive
allocations: because each individual's tail probabilities \(p_k(D)\)
decrease in \(k\), the vouchers it removes are those least likely to
be used. The cap chosen on the wrong prior, however, is itself too
large, because the fitted prior expects small requests and a cheap
price of information. As \(\varepsilon\) decreases to \(-4\), the
best fitted cap grows from \(8\) to \(10.5\) on average when
\(B_2=B_1\), and from \(14\) to the full \(20\) when \(B_2=1.5B_1\),
where the voucher price then degrades like the oracle. Reducing the cap restores
the limit on excessive allocations. The completion bonus has no cap:
the fitted population requests too little at every weight, so
\(\widehat\lambda\) rises above \(\lambda^\star\), where its
recruitment is sensitive (\cref{sec:exp-tuning}), and individuals
request vouchers up to their largest useful quantity
(\cref{ex:intro-linear}).

\paragraph{Overestimation (\(\varepsilon>0\)).}
For the oracle and the voucher price, the threshold
\(\widehat\theta\) is now set too high. Individuals receive too few
vouchers, leaving part of the voucher budget unused, and under the
voucher price each contract also reserves more rewards, so fewer
individuals are reached; a smaller cap only adds to the shortfall.
The completion bonus, by contrast, has a limit on how conservative it
can be: \(\widehat\lambda\) falls but cannot go below \(0\), so the
menu is never more conservative than the all-or-nothing menu, and
recruitment stops changing once this menu is reached. At \(B_1=5n\),
its recruitment falls from \(0.95\) to \(0.86\) by \(\varepsilon=2\)
and then stays there.

In short, a smaller cap insures the voucher price against
underestimating recruitment, at the price of lower recruitment when
the prior is accurate.

\section{Conclusion}
\label{sec:conclusion}

We formulated public-health referral allocation as a mechanism-design
problem with private referral prospects and separate voucher and
reward budgets. Our results show when simple contracts can recover
much of the recruitment attainable with full knowledge of
participants' prospects. The voucher-price mechanism asymptotically
matches the oracle when the reward budget is sufficiently large
relative to the voucher budget. The completion-bonus mechanism needs
no such surplus and achieves a tight asymptotic guarantee of
\(1-1/e\) for log-concave types. For unrestricted types and equal budgets, no fixed menu
guarantees a constant fraction of the oracle's recruitment. Simulations
based on HIV transmission network data illustrate the mechanisms'
different strengths across budget regimes and their different
responses to errors in the population prior.

Future work could extend the mechanisms to multiple recruitment
rounds, updating the population prior as referrals are observed,
building on \citet{pan2026adaptive}. Another direction is to combine
predictions based on observable characteristics with participants'
private knowledge. Finally, rewards could reflect programme
priorities, such as reaching first-time testers
\citep{zhou2022monetary}, so that allocations account for whom
participants recruit as well as how many.

\paragraph{Funding Acknowledgement.}
Avrim Blum, \'Eva Tardos, and Milind Tambe are supported by ONR MURI grant N00014-24-1-2742.
\paragraph{AI Disclosure.}
Generative AI tools, including GPT-5.6, GPT-6, Claude Opus 5.5, and Claude Fable 5.2, were used to assist with mathematical proofs, coding, and manuscript writing. The authors have reviewed and verified the AI-assisted content and take full responsibility for the accuracy of the manuscript.
\bibliographystyle{plainnat}
\bibliography{references}

\clearpage
\appendix
\etocdepthtag.toc{appendix}
\etocsettagdepth{mainmatter}{none}
\etocsettagdepth{appendix}{subsection}
\renewcommand{\contentsname}{Appendix Contents}
\tableofcontents
\section{Omitted Details for \texorpdfstring{\cref{sec:preliminaries}}{Section \getrefnumber{sec:preliminaries}}: Preliminaries}
\label{app:preliminaries}

\subsection{Menu Equivalence for \texorpdfstring{\cref{sec:mechanisms}}{Section \getrefnumber{sec:mechanisms}}}
\label{app:elicitation}

We prove the equivalence stated in \cref{sec:mechanisms} between
menu-based mechanisms and IR, IC direct-revelation mechanisms. Both
formulations use the same admissible contracts and a fixed rule that
depends only on the current individual's report or choice. All rules
and selections are measurable, with randomisation independent across
individuals and of capacities. The notation \(\mathcal M\) includes
the selection rule.

In a direct-revelation mechanism, a type \(D\) reports
\(\widetilde D\in\Delta(\Omega)\) and receives a contract drawn
from a distribution \(\Gamma(\widetilde D)\) over admissible
contracts. Its expected utility is
\[
 U(\widetilde D;D)
 :=\mathbb E_{C\sim\Gamma(\widetilde D)}[U(C;D)].
\]
A deterministic mechanism assigns a single contract
\(C(\widetilde D)\) to each report.

\begin{definition}[Individual rationality and incentive compatibility]
A mechanism is \emph{individually rational} (IR) if
\(U(D;D)\ge0\) for every type \(D\), with outside-option
utility normalised to zero. It is \emph{incentive-compatible} (IC) if
\(U(D;D)\ge U(\widetilde D;D)\) for every pair of types
\(D,\widetilde D\).
\end{definition}

\begin{proposition}[Menu and direct-revelation equivalence]
\label{prop:menu-elicitation}
Every deterministic IR and IC direct-revelation mechanism over
admissible contracts is implementable by an admissible fixed menu
under favourable tie-breaking. Conversely, every admissible fixed
menu with favourable tie-breaking induces an IR and IC
direct-revelation mechanism, possibly randomised.
The implementations preserve each type's contract outcomes,
resource reservations, and expected recruitment.
\end{proposition}

\begin{proof}
Given a deterministic IC direct-revelation mechanism, offer its range
\(\mathcal M=\{C(\widetilde D):\widetilde D\in\Delta(\Omega)\}\).
IC says exactly that \(C(D)\) maximises type \(D\)'s utility over this
menu, so favourable tie-breaking can select it. Maximum payments are
bounded by that of \(C(\delta_m)\): the type \(\delta_m\) completes every
quantity, so its utility from a contract is the contract's maximum
payment, and it weakly prefers its truthful contract. The selected
contracts, reservations, and stopping rule are unchanged.

Conversely, given a menu, let \(\sigma(D)\) be its prescribed
distribution over type \(D\)'s best responses, and on report
\(\widetilde D\) assign \(\Gamma(\widetilde D)=\sigma(\widetilde D)\).
Then
\[
 U(D;D)=\max_{C\in\mathcal M}U(C;D)
 \ge\mathbb E_{C\sim\sigma(\widetilde D)}[U(C;D)]
 =U(\widetilde D;D),
\]
which is IC. The selected outcomes coincide, and keeping the menu's
stopping thresholds preserves sequential recruitment.

IR holds in both formulations because payments are nonnegative and
utility is the expected payment.
\end{proof}

The converse direction produces a randomised direct mechanism only
because favourable tie-breaking may randomise among tied menu entries.
A direct mechanism that randomises over contracts that are not tied
would instead require menus whose entries are lotteries.

\subsection{Proof of \texorpdfstring{\cref{thm:relaxation-transfer}}{Theorem \getrefnumber{thm:relaxation-transfer}}: Relaxation and Sequential Recruitment}
\label{app:relaxation-transfer}

\begin{restatement}{\Cref{thm:relaxation-transfer} (restated)}
Every fixed menu of admissible contracts satisfies
\[
 \operatorname{Rec}(\mathcal M;B_1,B_2)
 \le\Phi(\mathcal M;B_1,B_2).
\]
If every offered contract has maximum payment at most \(H\), then
for \(B>0\),
\[
 \bigl(1-\epsilon_H(B)\bigr)\Phi(\mathcal M;B_1,B_2)
 \le\operatorname{Rec}(\mathcal M;B_1,B_2),
 \qquad
 \epsilon_H(B):=\sqrt{\frac HB}+\frac HB.
 \]
In particular, for fixed \(H\),
\(\operatorname{Rec}(\mathcal M;B_1,B_2)
=(1-o(1))\Phi(\mathcal M;B_1,B_2)\) as \(B\to\infty\).
\end{restatement}

\begin{proof}
Draw every individual's type, selection, and capacity in advance,
whether or not they are reached, and let \((Q_i,P_i,Z_i)\) be the
resulting quantity, maximum payment, and recruitment. Let \(I_i\)
indicate whether individual \(i\) is reached, and put
\(N=\sum_iI_i\). Since \(I_i\) depends only on preceding individuals,
it is independent of \((Q_i,P_i,Z_i)\). Hence expected recruitment is
\(r(\mathcal M)\mathbb E[N]\), and the budgets give
\(u(\mathcal M)\mathbb E[N]\le B_1\) and
\(e(\mathcal M)\mathbb E[N]\le B_2\). Together with \(N\le n\), this
proves the upper bound.

For the lower bound, write \(\mu_1=u(\mathcal M)\),
\(\mu_2=e(\mathcal M)\),
\[
 T=\min\{n,B_1/\mu_1,B_2/\mu_2\},\qquad \delta=H/B,
\]
and \(X_{i1}=Q_i\), \(X_{i2}=P_i\). For each finite budget \(B_j\), let
\(M_j=\sum_{i=1}^N(X_{ij}-\mu_j)\). Independence from preceding
individuals makes the cross terms vanish when its square is expanded,
giving
\[
 \mathbb E[M_j]=0,\qquad \mathbb E[M_j^2]
 =\operatorname{Var}(X_{1j})\mathbb E[N]
 \le H\mu_j\mathbb E[N]\le HB_j.
\]
If \(N<n\), then some budget \(j\) has fewer than \(H\) units left,
so \(\mu_jN+M_j>B_j-H\). Since \(\mu_j/B_j\le1/T\),
\[
 1-\frac NT
 \le1-\frac{\mu_jN}{B_j}
 <\frac H{B_j}+\frac{M_j}{B_j}
 \le\delta+\sum_j\frac{(M_j)_+}{B_j}.
\]
If \(N=n\), the same final bound holds because \(T\le n\). A centred
variable has \(\mathbb E[(M_j)_+]=\tfrac12\mathbb E|M_j|\), so taking
expectations and applying Cauchy--Schwarz gives
\[
 1-\frac{\mathbb E[N]}T
 \le\delta+\frac12\sum_j
       \frac{\sqrt{\mathbb E[M_j^2]}}{B_j}
 \le\delta+\sqrt\delta.
\]
Multiplying by \(r(\mathcal M)T=\Phi(\mathcal M)\) proves the claim.
\end{proof}

\section{Omitted Details for \texorpdfstring{\cref{sec:recruitment-maximizing}}{Section \getrefnumber{sec:recruitment-maximizing}}: Recruitment-Maximising Mechanisms}
\label{app:recruitment-maximizing}

\subsection{Four Necessary Contracts with Strict Preferences}
\label{app:menu-example}

We verify \cref{ex:four-contracts} and prove that four contracts are
necessary.
The expected-payment matrix, with types indexing rows and contracts
indexing columns, is
\[
 \begin{pmatrix}
 1+\varepsilon & 1 & 5/8+\varepsilon/2 & 1\\
 1+\varepsilon & 17/16+\varepsilon/2 & 15/16+3\varepsilon/4 & 17/16\\
 1+\varepsilon & 9/8+\varepsilon & 5/4+\varepsilon & 5/4\\
 1+\varepsilon & 9/8+\varepsilon & 5/4+\varepsilon & 2
 \end{pmatrix}.
\]
For \(0<\varepsilon<1/64\), every diagonal entry is a strict
row maximum. Hence each type \(D_i\) uniquely chooses \(C_i\).
All four contracts are admissible.

For this proof, write \(R=5r\), \(Q=5u\), and \(E=5e\) for
expected total recruitment, voucher use, and earmarked rewards if
all five individuals are reached. The displayed menu gives
\[
 R=2\cdot\tfrac12+\tfrac34+1+2=\tfrac{19}4,
 \qquad Q=6,
 \qquad E=2(1+\varepsilon)+(9/8+\varepsilon)
             +(5/4+\varepsilon)+2=\tfrac{51}8+4\varepsilon.
\]
Its relaxed recruitment is therefore
\[
 \Phi=R\min\left\{1,\frac6Q,\frac{B_2}{E}\right\}
 =\frac{19}4.
\]

\paragraph{Optimality and unique quantities.}
For each of \(D_1,D_2,D_3\), the unique maximiser of
\(v_D(q)-3q/4\) over \(q\in\{1,2\}\) is \(q=1\), because
its second-voucher marginal recruitment is less than \(3/4\).
For \(D_4\), the unique maximiser is \(q=2\). Weighting the four
maximum values by \((2,1,1,1)\) gives \(1/4\), so every menu and
selection rule satisfy
\[
 R\le\frac14+\frac34Q.
\]
If \(Q\le6\), then \(\Phi\le R\le19/4\). If \(Q\ge6\), then
\[
 \Phi\le\frac{6R}{Q}
 \le\frac92+\frac{3}{2Q}
 \le\frac{19}4.
\]
This proves optimality. Equality forces \(Q=6\) and equality in
the pointwise maximisation for every type. Thus every optimal
selection assigns one voucher to \(D_1,D_2,D_3\) and two vouchers
to \(D_4\), almost surely. It gives \(R=19/4\), and therefore
also requires \(E\le B_2\).

\paragraph{At least four contracts are necessary.}
Suppose an optimal menu contains at most three contracts.
At least one must allocate two vouchers, so at most two allocate
one voucher. For each of \(D_1,D_2,D_3\), select a one-voucher
best response having the smallest maximum reward among that type's
one-voucher best responses. This resolves any favourable
randomisation without changing quantities, incentives, or recruitment,
and cannot increase earmarked rewards. Denote the resulting expected
total earmark by \(E'\), so \(E'\le E\).

Any contract selected by \(D_4\) has maximum reward at least \(2\).
Nonnegativity of payments means the first three types can obtain
at least twice their probability of capacity two by choosing it.
Consequently their utilities satisfy
\[
 U_1\ge1,\qquad U_2\ge17/16,\qquad U_3\ge5/4.
\]
Every selected one-voucher contract has maximum reward at least its
selecting type's utility. Two of the first three types must share a
one-voucher contract. The three possible pairs give the following
lower bounds, including the other type's reward and the two reward
units required by \(D_4\):
\[
 \begin{array}{ccl}
 \text{shared types}& &\text{lower bound on }E'\\[2pt]
 D_1,D_2&&3\cdot\frac{17}{16}+\frac54+2=\frac{103}{16},\\[2pt]
 D_1,D_3&&3\cdot\frac54+\frac{17}{16}+2=\frac{109}{16},\\[2pt]
 D_2,D_3&&2\cdot\frac54+2\cdot1+2=\frac{13}{2}.
 \end{array}
\]
Thus
\[
 E\ge E'\ge\frac{103}{16}
 >\frac{51}8+4\varepsilon=B_2,
\]
contradicting optimality. Every relaxed-optimal menu therefore
requires at least four contracts, and the displayed menu attains
that minimum, exceeding \(|\Omega|=3\).

For completeness, the lower bound \(103/16\) is attainable when
implementing the same quantities with three contracts: use a
one-voucher flat payment \(17/16\), a one-voucher completion payment
\(5/4\), and a two-voucher completion payment \(2\), assigning
the first to \(D_1,D_2\), the second to \(D_3\), and the last to
\(D_4\), with favourable tie-breaking.

The unique quantity assignment and strict preferences do not imply
unique optimal payment schedules. For example, replacing
\(\varepsilon\) in all four contracts by any
\(\varepsilon'\in(0,\varepsilon)\), while keeping the instance's
budgets fixed, preserves strict preferences and recruitment and
reduces earmarked rewards. It yields another optimal menu.

\begin{proposition}[Strict choices do not identify a unique optimal menu]
\label{prop:strict-nonuniqueness}
Suppose the population has finite type support and every type has a
unique utility-maximising contract in a finite menu. Then there is a
continuum of distinct menus inducing exactly the same selected
quantity and maximum reward for every type. This remains true after
discarding contracts selected by no type. In particular, if the
original menu maximises relaxed or sequential recruitment, it is not
the unique recruitment-maximising menu of payment schedules.
\end{proposition}

\begin{proof}
Fix a contract \(C=(q,\tau)\) selected by at least one type, and put
\(h=\tau(q)\). Keep \(q\), \(h\), and every other contract fixed.
If \(\tau\) is not the constant rule \(h\), let
\(\widehat\tau(z)=h\) for all \(z\). Otherwise, let
\(\widehat\tau(z)=h\mathbf 1\{z\ge q\}\).
In either case, \(\widehat\tau\) is admissible, has maximum reward
\(h\), and differs from \(\tau\), since \(q\ge1\) and \(h\ge q>0\).
For \(\varepsilon\in(0,1)\), replace \(\tau\) by
\[
 \tau^\varepsilon=(1-\varepsilon)\tau+
 \varepsilon\widehat\tau.
\]
Admissibility and the maximum reward are preserved. All best-response
inequalities are strict, and there are finitely many of them.
Consequently, for all sufficiently small positive \(\varepsilon\),
every type continues to select the same contract, with \(C\) replaced
by \((q,\tau^\varepsilon)\). These are distinct menus as
\(\varepsilon\) varies, and the modified contract remains selected.

The quantity and earmarked reward selected by each type are unchanged,
so \(r,u,e\) and relaxed recruitment are unchanged. For the sequential
process, couple the type and capacity draws across the menus. The
maximum offered quantity and earmark, every allocation and earmark,
and therefore the stopping time and referral count are identical on
every sample path. Sequential recruitment is thus unchanged as well.
\end{proof}

\subsection{Proof of \texorpdfstring{\cref{thm:menu-hardness}}{Theorem \getrefnumber{thm:menu-hardness}}: Recruitment-Maximising Menus}
\label{app:menu-hardness}

We give a reduction from \textsc{Vertex Cover} that uses only positive
quantities and capacities. The argument permits arbitrary admissible
payment schedules and favourable randomisation among best responses.

\paragraph{Construction.}
Let $G=(\{1,\ldots,N\},\mathcal E)$ be a graph with $N\ge2$,
$\ell=|\mathcal E|$, and cover target $0\le k\le N$.
Set $m=9$, $L=96$, and $J=N+1$. We specify a type by its
\emph{scaled tails} $a_r=L\Pr(R\ge r)$, for $1\le r\le9$.
Define
\[
 g=(96,18,17,16,15,14,13,12,6),\qquad
 z_i=(0,1,i,i^2,i^3,i^4,-s_i,0,0),\qquad
 s_i=\sum_{r=0}^4i^r.
\]
For distinct vertex labels $i,j$, let
\[
 P_{ij}(y)=\frac{2(y-i)}{j-i}-2N(y-i)^2(y-j)^2.
\]
Let $Z=\max_i\|z_i\|_\infty$, and let $A\ge1$ be the largest
absolute coefficient of these ordered-pair polynomials. Set
\[
 \varepsilon=\min\{1/Z,1/(8A),1\},\qquad
 \rho=\varepsilon^2/2,\qquad
 M=\left\lceil\frac{12LN}{\rho}\right\rceil.
\]
There is one guard type, one type for each vertex, and one type for
each edge. Their scaled tails and prior weights are as follows:
\begin{center}
\small
\begin{tabular}{@{}lccc@{}}
\toprule
Type & $a_1$ & $a_r$, $2\le r\le9$ & Weight\\
\midrule
Guard & $L$ & $g_r$ & $M$\\
Vertex $i$ & $L$ & $3g_r+\varepsilon z_{i,r}$ & $1$\\
Edge $\{i,j\}$ & $L$ & $5g_r+\varepsilon(z_{i,r}+z_{j,r})/2-\rho$ & $J$\\
\bottomrule
\end{tabular}
\end{center}
There are $n=M+N+J\ell$ initial individuals, and a type of weight
$w$ has probability $w/n$. These are valid distributions supported
on $\{1,\ldots,9\}$. Indeed, all first tails equal $L$;
vertex and edge second tails are at most $55$ and $91$, respectively.
Each feature perturbation has magnitude at most one, whereas
successive base tails before coordinate $8$ differ by at least
three. The last two tails are $(36,18)$ for a vertex and
$(60-\rho,30-\rho)$ for an edge. Thus the tails are nonincreasing
and nonnegative. Probabilities are their successive differences,
divided by $L$.

\paragraph{An auxiliary objective.}
Write a contract through the nonnegative increments
$x_0=\tau(0)$ and $x_r=\tau(r)-\tau(r-1)$ for $1\le r\le9$.
The increments vanish for $r>q$, and
$h=\sum_{r=0}^9x_r\ge q$ is its earmarked reward.
For a type $t$, its scaled utility is
\[
 U_t(x)=Lx_0+\sum_{r=1}^9a_{t,r}x_r.
\]
In particular, put $b=x_0+x_1=\tau(1)$.
For this proof only, let $R,Q,E$ denote the expected \emph{totals}
of recruitment, vouchers, and earmarked rewards if all $n$ individuals
are reached; thus $R=nr(\mathcal M)$, $Q=nu(\mathcal M)$, and
$E=ne(\mathcal M)$.
Consider
\[
 W=LR-18Q-E,\qquad
 \overline W=\sum_t w_t\max_{1\le q\le9}
 \left\{\sum_{r=1}^q a_{t,r}-19q\right\}.
\]
Since $h\ge q$, $\overline W$ is a pointwise upper bound on $W$.
Comparing each tail with the marginal cost $19$ gives
\begin{center}
\begin{tabular}{@{}lll@{}}
\toprule
Type & Pointwise best quantity & Loss at every other quantity\\
\midrule
Guard & $1$ & At least $1$\\
Vertex & $8$ & At least $1$; exactly $1$ at $q=h=9$\\
Edge & $9$ & At least $11-\rho$\\
\bottomrule
\end{tabular}
\end{center}
Excess payment $h-q$ adds that amount to the loss. In particular,
every individual contribution to $\overline W-W$ is nonnegative,
and direct summation gives
\[
 \overline W=77M+259N+(480-8\rho)J\ell.
\]

\begin{lemma}[Implementing a vertex cover]
\label{lem:hardness-cover-completeness}
Every vertex cover $C$ admits a menu with earmarked rewards at most
$9$ such that
\[
 W=\overline W-|C|,\qquad
 Q=E=M+8N+|C|+9J\ell.
\]
\end{lemma}
\begin{proof}
First offer the completion contracts with quantities and rewards
$(q,h)=(1,1)$ and $(8,8)$. The guard's scaled utility from either
is $L$; every vertex's utility from the latter is $3L$.

For $P(y)=\sum_{r=0}^4 A_ry^r$, where $P=0$ or $P=P_{ij}$,
construct a quantity-$9$ contract by setting
\[
 x_{r+2}=\frac14+\varepsilon A_r\quad(0\le r\le4),\qquad
 x_7=\frac14,\qquad x_0=x_1=0.
\]
Let $s=\sum_{r=2}^7x_r$ and
$u=\frac1{12}\sum_{r=2}^7g_rx_r$, and set
\[
 x_8=7+s-2u,\qquad x_9=2+2(u-s).
\]
All these increments are nonnegative. In fact,
$x_2,\ldots,x_7\in[1/8,3/8]$, so $s\le9/4$ and
$s\le u\le3s/2$. Consequently $x_9\ge2$ and
$x_8\ge7-2s\ge5/2$. Their sum is $h=9$, so the contract is
admissible. It satisfies
\[
 U_g=12u+12x_8+6x_9=L,\qquad
 z_i\cdot x=\varepsilon P(i).
\]
Thus the guard, vertex, and edge utilities from this contract are
\begin{align*}
 U_g&=L,\\
 U_i&=3L+\varepsilon^2P(i),\\
 U_{ij}&=5L+\varepsilon^2\frac{P(i)+P(j)}2-9\rho.
\end{align*}

Offer the neutral polynomial $P=0$. For each edge $\{i,j\}$,
choose an endpoint $j\in C$, orient the edge as $(i,j)$, and offer
the contract for $P_{ij}$. At vertex labels,
$P_{ij}(i)=0$, $P_{ij}(j)=2$, and
$P_{ij}(r)\le-2$ for $r\notin\{i,j\}$: at such an integer label
the squared product is at least one and the linear term is at most
$2(N-1)$.

Vertices outside $C$ can therefore choose quantity $8$ as a best
response. Vertices in $C$ choose a utility-maximising quantity-$9$
contract, using the neutral contract when needed to resolve a tie.
An edge's own polynomial contract gives it utility
$5L+\varepsilon^2-9\rho$, which exceeds its utility
$5L-8\rho$ from the quantity-$8$ contract because
$\varepsilon^2-\rho>0$. It also exceeds $L$, its utility from
the quantity-$1$ contract. Every edge therefore chooses $9$.
The guard chooses $1$. All selected contracts have $h=q$;
precisely the vertices in $C$ incur a one-unit auxiliary loss.
\end{proof}

\begin{lemma}[Extracting a vertex cover]
\label{lem:hardness-cover-soundness}
Every menu with loss $D:=\overline W-W\le N$ determines a vertex
cover of size at most $D$, including when ties are randomised.
\end{lemma}
\begin{proof}
Put $\eta=LN/M\le\rho/12$.

\emph{The guard bounds every offered contract.}
For every $q\ge1$,
$\sum_{r=1}^qg_r\le96+18(q-1)$. Hence the guard's loss under a
selected contract is at least
\[
 77-\sum_{r=1}^qg_r+18q+h\ge h-1.
\]
Since the guard has weight $M$, its mean selected terminal reward
is at most $1+N/M$. All its selected contracts attain the same
maximum utility, and $U_g(x)\le Lh$. It follows that
\begin{equation}
 U_g(x)\le L+\eta
 \quad\text{for every offered contract }x.
 \label{eq:hardness-guard-cap}
\end{equation}

\emph{Vertices receiving quantity $8$ have bounded utility.}
Let $S$ contain the vertices that select quantity $8$ with positive
probability. For any offered quantity-$8$ contract, put
$s=\sum_{r=0}^7x_r$. The guard coefficient at coordinate $8$ is
$12$, and every preceding coefficient is at least $13$. Thus
\[
 U_g\ge12h+s\ge96+s,\qquad s\le\eta.
\]
Using $b\ge0$ and $\varepsilon\|z_i\|_\infty\le1$, we obtain
\[
 U_i=3U_g-2Lb+\varepsilon z_i\cdot x
 \le3L+3\eta+s\le3L+4\eta.
\]
For $i\in S$, a quantity-$8$ contract is a best response, so this
also bounds $i$'s utility from every offered contract.

\emph{No edge has both endpoints in $S$.}
Suppose $\{i,j\}\in\mathcal E$ and $i,j\in S$. Every offered
quantity-$9$ contract satisfies
\begin{align*}
 U_{ij}
 &=\frac{U_i+U_j}{2}+2U_g-\rho h-(2L-\rho)b\\
 &\le5L+6\eta-9\rho
 <5L-8\rho.
\end{align*}
On the other hand, a quantity-$8$ contract is offered since
$i\in S$. All of this edge's utility coefficients up to coordinate
$8$, including the base-payment coefficient, are at least
$60-\rho$. Its utility from that contract is therefore at least
\[
 (60-\rho)h\ge8(60-\rho)=5L-8\rho.
\]
The edge never chooses quantity $9$, and so contributes at least
$J(11-\rho)>N$ to the loss, a contradiction.

Consequently $C=V\setminus S$ is a vertex cover. Every vertex in
$C$ receives a quantity other than $8$ with probability one and
incurs loss at least one. Therefore $|C|\le D$.
\end{proof}

Writing $\tau(G)$ for the minimum vertex-cover size, the two lemmas
give
\begin{equation}
 \max_{\mathcal M} W=\overline W-\tau(G).
 \label{eq:hardness-auxiliary-optimum}
\end{equation}
Indeed, if $D>N$, then already $D\ge\tau(G)$; otherwise apply
\cref{lem:hardness-cover-soundness}.

\paragraph{A gap for the expected-budget relaxation.}
Set
\[
 B=M+8N+k+9J\ell,\qquad T=\overline W-k,\qquad
 R_k=\frac{T+19B}{L},\qquad B_1=B_2=B.
\]
Here $T-1>0$. In a YES instance, extend a cover to size exactly
$k$. The menu from \cref{lem:hardness-cover-completeness} has
$Q=E=B$, $W=T$, and relaxed recruitment $R_k$.
In a NO instance, \cref{eq:hardness-auxiliary-optimum} implies
$W\le T-1$ for every menu. With
$\alpha=\min\{1,B/Q,B/E\}$,
\[
 L\Phi=\alpha W+18\alpha Q+\alpha E
 \le\max\{W,0\}+19B
 \le T-1+19B.
\]
Thus YES instances have relaxed optimum at least $R_k$, whereas
NO instances have relaxed optimum at most $R_k-1/L$.
This is a gap in expected total recruitment, not per-individual
recruitment.

\paragraph{Transfer to sequential recruitment.}
Keep the prior and all type distributions fixed, and multiply
$n,B_1,B_2$ by the integer
$s=16mL^2B$. Every menu's relaxed value is multiplied by $s$.
The YES menu has maximum payment $H=m=9$ and satisfies $R_k\le B$.
By \cref{thm:relaxation-transfer}, its sequential recruitment is
at least $sR_k-\sqrt{msB}-m$, and
$\sqrt{msB}+m\le s/(2L)$.
For a NO instance, the relaxation bounds every menu from above.
We therefore have the separating gap
\[
 \begin{array}{ll}
 \text{YES:}&\displaystyle
 \sup_{\mathcal M}\operatorname{Rec}(\mathcal M;sB,sB)
 \ge sR_k-\frac{s}{2L},\\[4pt]
 \text{NO:}&\displaystyle
 \sup_{\mathcal M}\operatorname{Rec}(\mathcal M;sB,sB)
 \le sR_k-\frac{s}{L}.
 \end{array}
\]
An algorithm returning an optimal menu and its selection rule would
also distinguish these cases: its relaxed value, computed from the
finite prior, exceeds $s(R_k-3/(4L))$ in a YES instance and is below
that threshold in a NO instance. This avoids any need to evaluate
the sequential stopping-time expectation exactly.

\paragraph{Encoding and strong hardness.}
The construction has $1+N+\ell$ types. Since
$Z=O(N^4)$ and $A=O(N^5)$, with polynomially bounded coefficient
denominators, $\varepsilon^{-1}=O(N^5)$,
$\rho^{-1}=O(N^{10})$, and $M=O(N^{11})$.
All rational numerators and denominators are polynomially bounded,
as are the integers $n,B,sn,sB$ (the last two are $O(N^{22})$).
Hence both reductions are polynomial even with these numerical
data encoded in unary, proving strong NP-hardness.

\Needspace{12\baselineskip}
\section{Omitted Proof for \texorpdfstring{\cref{sec:general}}{Section \getrefnumber{sec:general}}: Unrestricted Types}
\label{app:general}
\label{app:unrestricted-proof}

We give the full proof of \cref{prop:warmup-impossibility}.
The argument allows nondecreasing rewards, nonnegative base payments,
and favourable randomisation among tied contracts.

\begin{restatement}{\Cref{prop:warmup-impossibility} (restated)}
For unrestricted types, the best worst-case approximation ratio is
\(O(1/\sqrt m)\) in the large-budget limit. For every \(m\ge2\)
and integer \(B>m\), there is an instance with
\(n=B_1=B_2=B\) on which every admissible fixed menu satisfies
\[
 \frac{\operatorname{Rec}(\mathcal M;B,B)}{\mathrm{OPT}(B)}
 \le\frac{5B}{2\sqrt m\,(B-m)}.
\]
\end{restatement}

\begin{proof}
For \(2\le m\le6\), the stated bound exceeds one, so any instance
with positive recruitment suffices. Suppose \(m\ge7\), and put
\(t=m^{1/4}\). The population has three types: reliable recruiters \(H=\delta_m\) with probability
\(w_H=t^{-4}\); occasional recruiters
\(L=(1-p)\delta_0+p\delta_m\), where \(p=t^{-2}\), with probability
\(w_L=t^{-1}\); and zero-capacity individuals \(\delta_0\) with
probability \(w_0=1-w_H-w_L\).

\emph{Incentives.} For each type \(j\in\{H,L,0\}\), let
\(Q_j,P_j,b_j\) be the mean selected quantity, maximum payment, and
payment at zero referrals, averaging over tied choices. Since \(L\)
recruits either nobody or at least any offered quantity, a contract
with maximum payment \(P\) and payment \(b\) at zero gives \(L\)
utility \(pP+(1-p)b\); the zero type receives \(b\). Type \(L\) can
choose \(H\)'s contracts, and the zero type can choose \(L\)'s, so
\[
 pP_H\le pP_L+(1-p)b_L\le pP_L+(1-p)b_0\le pP_L+(1-p)P_0.
\]
Thus attracting reliable recruiters requires either large rewards for
occasional recruiters or base payments that also attract the
zero-capacity group.

\emph{Menu recruitment.} Put
\(s=(w_H+w_Lp)/(w_H+w_L)=1/(t^2-t+1)\). Then
\(w_H(1-s)=w_L(s-p)\), and a direct computation gives
\[
 sw_0-w_H(1-s)\frac{1-p}{p}=\frac{t^2-t-1}{t^4(t^2-t+1)}\ge0,
\]
since \(m\ge7\). Using \(Q_j\le P_j\) and the incentive inequality,
\[
\begin{aligned}
 r&=w_HQ_H+w_LpQ_L\le w_HP_H+w_LpP_L
  =s(w_HP_H+w_LP_L)+w_H(1-s)(P_H-P_L)\\
  &\le s(w_HP_H+w_LP_L)+w_H(1-s)\frac{1-p}{p}P_0
  \le s(w_HP_H+w_LP_L+w_0P_0)=se.
\end{aligned}
\]
Hence every menu recruits at most \(r\cdot B/e\le sB\).

\emph{Oracle recruitment.} The oracle gives \(m\) vouchers to \(H\)
and one to everyone else, so its mean voucher use and recruitment are
\(u=2-t^{-4}\) and \(r^\star=1+t^{-3}\). Since \(n=B\) and each
reached individual receives at least one voucher, it uses at least
\(B-m\) vouchers on every run. As reaching an individual is
independent of their type, its recruitment is at least
\(\rho(B-m)\), where \(\rho=r^\star/u\). Consequently,
\[
 \frac{\operatorname{Rec}(\mathcal M;B,B)}{\mathrm{OPT}(B)}
 \le\frac{B}{B-m}\cdot\frac{s}{\rho}
 =\frac{B}{B-m}\cdot\frac{(t+1)(2t^4-1)}{t(t^3+1)^2}
 \le\frac{5B}{2\sqrt m\,(B-m)}.
\]
For the last inequality, clearing denominators leaves
\(t^3(t^3-4t^2+10)+2t^2+2t+5>0\), which holds since
\(t^3-4t^2+10\ge14/27\) for \(t\ge0\). The exact factor \(s/\rho\) is
\((2+O(m^{-1/4}))/\sqrt m\).
\end{proof}

\Needspace{12\baselineskip}
\section{Omitted Proofs for \texorpdfstring{\cref{sec:restricted}}{Section \getrefnumber{sec:restricted}}: Approximation Guarantees}
\label{app:restricted}

This appendix proves the results of \cref{sec:restricted}. We first
fix notation and prove a precise version of
\cref{lem:shared-attainability} (\cref{app:menu-attainability}). We
then reduce \cref{thm:restricted-guarantees} to two lemmas about a
single type (\cref{app:restricted-proof}). Both lemmas rest on tail
estimates for log-concave types at a local maximum
(\cref{app:local-tail-bounds}), which give the pointwise comparison
almost immediately (\cref{app:pointwise-comparison}) and are the main
input to the comparison of averages (\cref{app:starting-density}).

Throughout, for a type \(D\) we write \(d_j=D(j)\),
\(p_j=\Pr_D(R\ge j)\), with \(p_0=1\) and \(p_{m+1}=0\), and
\(v(q)=\sum_{j=1}^q p_j\), suppressing the dependence on \(D\) when it
is clear. We call \(\delta_0\) the \emph{zero type} and every other
type \emph{positive}, so a positive type has \(p_1>0\). From \cref{app:restricted-proof} on, types are log-concave
(\cref{def:lc-types}), and we write
\[
 \alpha:=1-\frac1e,\qquad \beta_e:=\frac1{2e},
 \qquad\text{so that}\qquad \alpha=1-2\beta_e.
\]

\subsection{Proof of \texorpdfstring{\cref{lem:shared-attainability}}{Proposition \getrefnumber{lem:shared-attainability}}: Attainable Mean Quantities}
\label{app:menu-attainability}

Write \(H_\lambda(q):=(1-\lambda)qp_q+\lambda v(q)\) for type \(D\)'s
expected payment from quantity \(q\) under \(\mathcal C_\lambda\), and
let
\[
 \mathcal K_\lambda(D):=
 \argmax_{\substack{q\in\mathcal Q\\k_0(D)\le q\le\bar b(D)}}H_\lambda(q)
\]
be the set of best responses in \(\{k_0(D),\ldots,\bar b(D)\}\), and put
\[
 k_\lambda^-(D):=\min\mathcal K_\lambda(D),\qquad
 k_\lambda^+(D):=\max\mathcal K_\lambda(D),\qquad
 u_\lambda^\pm:=\mathbb E[k_\lambda^\pm(D)].
\]
The following is the precise form of \cref{lem:shared-attainability}.

\begin{restatement}{\Cref{lem:shared-attainability} (precise form)}
For each type \(D\), \(k_\lambda^-(D)\) and \(k_\lambda^+(D)\) are
nondecreasing in \(\lambda\), and for a positive type and \(\lambda<1\),
\(\mathcal K_\lambda(D)\) is the set of all best responses under
\(\mathcal C_\lambda\). For each \(u\in[u_0,u_{\max}]\), the set
\(\{\lambda:u_\lambda^+\ge u\}\) is an interval \([\lambda_u,1]\), and
with favourable tie-breaking \(\mathcal C_{\lambda_u}\) admits a
selection from \(\mathcal K_{\lambda_u}(D)\) with mean exactly \(u\).
\end{restatement}

\begin{proof}
\emph{Monotone choices.} For the zero type, every quantity has utility
zero and \(\mathcal K_\lambda(D)=\{1\}\). Let \(D\) be positive. At
\(\lambda=1\), \(H_1=v\) is maximised over
\([k_0(D),\bar b(D)]\) only at \(\bar b(D)\), so
\(\mathcal K_1(D)=\{\bar b(D)\}\). For \(\lambda<1\), a quantity
\(q>b_+(D)\) earns the same linear reward as \(b_+(D)\) but forfeits
its positive completion bonus, so every best response lies in
\(\{1,\ldots,b_+(D)\}\), where \(v\) is strictly increasing. Dividing
by \(1-\lambda\), best responses maximise \(qp_q+sv(q)\) with
\(s=\lambda/(1-\lambda)\). If \(k\) is optimal at \(s\) and \(k'\) at
\(s'>s\), adding the two optimality inequalities gives
\((s'-s)(v(k')-v(k))\ge0\), so \(k'\ge k\). Thus every best response
at a weight \(\lambda<1\) is at least every best response at a smaller
weight, in particular at least \(k_0(D)\), and at most
\(b_+(D)=\bar b(D)\). Hence, for \(\lambda<1\), \(\mathcal K_\lambda(D)\)
is the full set of best responses, and \(k_\lambda^\pm(D)\) are
nondecreasing on \([0,1]\).

\emph{Filling the jumps.} Note that \(u_0^-=u_0\) and \(u_1^+=u_{\max}\). Each \(H_\lambda(q)\) is
affine in \(\lambda\) and there are finitely many quantities, so for
\(\mu\) close enough to \(\lambda\), every best response at \(\mu\) is
also one at \(\lambda\); by monotonicity, it is \(k_\lambda^-(D)\) when
\(\mu<\lambda\) and \(k_\lambda^+(D)\) when \(\mu>\lambda\). By bounded
convergence, \(\lambda\mapsto u_\lambda^+\) is nondecreasing and
right-continuous, with left limits \(u_\lambda^-\). Hence, for
\(u\in[u_0,u_{\max}]\), the set \(\{\lambda:u_\lambda^+\ge u\}\) is an
interval \([\lambda_u,1]\), and
\(u_{\lambda_u}^-\le u\le u_{\lambda_u}^+\): if \(\lambda_u>0\), because
\(u_\mu^+<u\) for all \(\mu<\lambda_u\), and otherwise because
\(u_0^-=u_0\). Select \(k_{\lambda_u}^+(D)\) with probability
\((u-u_{\lambda_u}^-)/(u_{\lambda_u}^+-u_{\lambda_u}^-)\) and
\(k_{\lambda_u}^-(D)\) otherwise, independently across individuals,
using \(k_{\lambda_u}^-(D)\) if the two means coincide. This mixes only
tied best responses and attains mean \(u\).
\end{proof}

\paragraph{Finite populations.} For a population with finite support,
each \(H_\lambda(q)\) is affine in \(\lambda\), so \(u_\lambda^+\) can
change only at \(0\), \(1\), and the weights in \([0,1]\) at which two
of a type's utility lines intersect. Since \(\lambda\mapsto u_\lambda^+\)
is right-continuous, \(\lambda^\star=\lambda_{u^\star}\) is one of these
candidates. A prior with \(N\) types has \(O(Nm^2)\) candidates, and
\(u_\lambda^\pm\) can be evaluated at each in \(O(Nm)\) time, so
\(\lambda^\star\) and the mixing probability in the proof above are
computable in polynomial time.

\subsection{Proof of \texorpdfstring{\cref{thm:restricted-guarantees}}{Theorem \getrefnumber{thm:restricted-guarantees}}: Reduction to Two Lemmas}
\label{app:restricted-proof}

Let \(r_0:=\mathbb E[v_D(k_0(D))]\) denote the mean recruitment under
all-or-nothing payments. The proof combines two comparisons for a
single type. The first compares recruitment at equal mean
allocations; the second controls recruitment per voucher when
all-or-nothing payments already allocate at least as many vouchers as
the comparison rule.

\begin{lemma}[Pointwise comparison of tail sums]
\label{lem:family-matching}
\label{lem:uniform-matching}
\label{lem:matching}
Let \(D\) be a log-concave type. Take \(1\le k\le\bar b(D)\), let \(\lambda\in[0,1]\), and assume
\(H_\lambda(k)\ge H_\lambda(k-1),H_\lambda(k+1)\), omitting
neighbours outside \(\mathcal Q\). Then, for every \(q\in\mathcal Q\),
\begin{equation}
 v(k)-\tfrac7{10}v(q)\ge c_\lambda(k-q),\qquad
 c_\lambda:=\tfrac{3(1-\lambda)}{10}.
 \label{eq:family-matching}
\end{equation}
\end{lemma}

The hypotheses of \cref{lem:family-matching} hold for every
\(k\in\mathcal K_\lambda(D)\). For the zero type, \(k=1\) and the lemma
is immediate. For a positive type and \(\lambda<1\), \(\mathcal K_\lambda(D)\)
is the set of all best responses, so \(k\) maximises \(H_\lambda\)
globally and hence locally. For \(\lambda=1\),
\(\mathcal K_1(D)=\{\bar b(D)\}\), and \(H_1=v\) is nondecreasing and
constant beyond \(\bar b(D)\). The correction \(c_\lambda(k-q)\)
averages to zero when the two mean quantities agree.

\begin{lemma}[Comparison of averages]
\label{lem:starting-density}
\label{lem:uniform-density}
Let \(Q\in\mathcal Q\) be any random integer jointly distributed
with \(D\sim\mathfrak D\), and put
\(u:=\mathbb E[Q]\), \(r:=\mathbb E[v_D(Q)]\).
If \(u\le u_0\), then \(r_0/u_0\ge\alpha r/u\), where
\(\alpha=1-1/e\).
\end{lemma}

\begin{proof}[Proof of \cref{thm:restricted-guarantees}]
Write \(u^\star=u(A^\star)\), \(r^\star=r(A^\star)\), and
\(\rho^\star=r^\star/u^\star\) for the oracle's mean allocation,
mean recruitment, and recruitment per voucher. If \(u_0\ge u^\star\),
select \(k_0(D)\) under \(\mathcal C_0\); \cref{lem:starting-density}
with \(Q=A^\star(D)\) gives \(r_0/u_0\ge\alpha\rho^\star\). Otherwise,
\cref{lem:shared-attainability} supplies a selection under
\(\mathcal C_{\lambda^\star}\) with mean \(u=u^\star\), supported on
\(\mathcal K_{\lambda^\star}(D)\). Averaging \eqref{eq:family-matching}
with \(q=A^\star(D)\) cancels the quantity difference and gives
\(r\ge\frac7{10}r^\star\ge\alpha r^\star\).

In either branch, the selection has \(u\ge u^\star\),
\(r/u\ge\alpha\rho^\star\), and \(e=u\), since every contract in
\(\mathcal C_\lambda\) earmarks exactly its quantity. Consequently,
\[
 \Phi(\mathcal M^\star)
 \ge\frac r u\min\{nu,B\}
 \ge\alpha\rho^\star\min\{nu^\star,B\}=\alpha F(B).
\]
Maximum payments are at most \(m\), so
\cref{thm:relaxation-transfer} yields
\(\operatorname{Rec}(\mathcal M^\star;B_1,B_2)
\ge(\alpha-o(1))F(B)\ge(\alpha-o(1))\mathrm{OPT}(B)\).
\end{proof}

\subsection{Tail Estimates at a Local Maximum}
\label{app:local-tail-bounds}

Log-concave masses satisfy \(d_jd_{j+2}\le d_{j+1}^2\) on an integer
interval. Hence
\(p_j/d_j=\sum_{\ell\ge0}d_{j+\ell}/d_j\) is nonincreasing on the
support, and the tail ratio \(p_{j+1}/p_j=1-d_j/p_j\) is nonincreasing
wherever \(p_j>0\); before the support it equals one.

Both comparison lemmas concern a quantity \(k\) that maximises
\(H_\lambda(q)=(1-\lambda)qp_q+\lambda v(q)\) locally. Writing
\(\sigma=1-\lambda\), the identity
\(H_\lambda(j)-H_\lambda(j-1)=p_j-\sigma(j-1)d_{j-1}\) shows that this
local optimality is equivalent to
\begin{equation}
 p_k\ge\sigma(k-1)d_{k-1},\qquad
 p_{k+1}\le\sigma k d_k,
 \label{eq:adjacent-tail-conditions}
\end{equation}
where the first condition is omitted when \(k=1\). At \(\lambda=0\),
these are the conditions for a local maximum of \(jp_j\); in particular,
they hold with \(\sigma=1\) at \(k_0(D)\).

\begin{lemma}[Tail estimates]
\label{lem:tail-estimates}
Let \(D\) be a positive log-concave type, \(0<\sigma\le1\), and
\(1\le k\le b_+(D)\) satisfy \eqref{eq:adjacent-tail-conditions}.
Put \(s:=\sigma(k-1)/(1+\sigma(k-1))\). Then
\begin{enumerate}[(i)]
\setlength{\itemsep}{1pt}
\item \(p_k\ge s^k\) and \(v(k)\ge\sigma(k-1)(1-s^k)\);
\item for \(0\le\ell\le m-k\),
\(
 v(k+\ell)-v(k)\le\sigma kp_k\bigl(1-e^{-\ell/(\sigma k)}\bigr)
 \le\sqrt{\sigma v(k)\ell/2}.
\)
\end{enumerate}
\end{lemma}

\begin{proof}
Substituting \(d_j=p_j-p_{j+1}\) into
\eqref{eq:adjacent-tail-conditions} gives \(p_k/p_{k-1}\ge s\) when
\(k\ge2\), and \(p_{k+1}/p_k\le r:=\sigma k/(1+\sigma k)\). Since the
tail ratios are nonincreasing, \(p_j\ge s^j\) for \(j\le k\), and
\(p_{k+j}\le p_kr^j\) for \(j\ge0\). Summing the first bound gives
\(v(k)\ge s(1-s^k)/(1-s)=\sigma(k-1)(1-s^k)\); for \(k=1\), part (i)
is trivial since \(s=0\). Summing the second gives
\(v(k+\ell)-v(k)\le\sigma kp_k(1-r^\ell)\), and
\(r^\ell=(1+1/(\sigma k))^{-\ell}\ge e^{-\ell/(\sigma k)}\).
Finally, by Cauchy--Schwarz,
\((1-e^{-x})^2=(\int_0^x e^{-y}\,dy)^2\le x/2\); with
\(x=\ell/(\sigma k)\) and \(kp_k^2\le kp_k\le v(k)\), this gives the
last bound.
\end{proof}

\begin{lemma}[Comparison at a local maximum]
\label{lem:completion-comparison}
Under the hypotheses of \cref{lem:tail-estimates}, every
\(q\in\mathcal Q\) satisfies
\begin{equation}
 v(k)-\alpha_0 v(q)\ge c(k-q)
 \label{eq:local-comparison}
\end{equation}
in each of the cases
(a) \(\sigma=1\) and \((\alpha_0,c)=(3/4,3/8)\);
(b) \((\alpha_0,c)=(7/10,3\sigma/10)\).
In particular, every log-concave type satisfies
\begin{equation}
 v(k_0)-\tfrac34v(q)\ge\tfrac38(k_0-q)
 \qquad\bigl(k_0=k_0(D),\ q\in\mathcal Q\bigr).
 \label{eq:completion-matching}
\end{equation}
\end{lemma}

\begin{proof}
In both cases \(\alpha_0\le3/4\) and \(c\ge3\alpha_0\sigma/8\).
For \(q=k+\ell>k\), \cref{lem:tail-estimates}(ii) and AM--GM give
\[
 v(q)-v(k)\le\sqrt{\sigma v(k)\ell/2}
 \le\frac{v(k)}3+\frac{3\sigma\ell}8,
\]
hence
\(v(k)-\alpha_0 v(q)\ge(1-4\alpha_0/3)v(k)-3\alpha_0\sigma\ell/8\ge-c\ell\).

For \(q<k\), tails are nonincreasing, so \(v(k)-v(q)\ge(k-q)p_k\) and
\[
 v(k)-\alpha_0 v(q)=(1-\alpha_0)v(k)+\alpha_0\bigl(v(k)-v(q)\bigr)
 \ge(1-\alpha_0)v(k)+\alpha_0(k-q)p_k.
\]
The difference between the right-hand side and \(c(k-q)\) is affine
in \(k-q\in[0,k-1]\) and nonnegative at \(k-q=0\). At \(k-q=k-1\),
\cref{lem:tail-estimates}(i) bounds it below by
\((k-1)[(1-\alpha_0)\sigma(1-s^k)+\alpha_0 s^k-c]\). In case~(b) the
bracket equals \(s^k(\alpha_0-(1-\alpha_0)\sigma)\ge0\). In case~(a) it
equals \(s^k/2-1/8\), and \(s^k=(1-1/k)^k\ge1/4\) for \(k\ge2\).

For \eqref{eq:completion-matching}, a positive type satisfies
\eqref{eq:adjacent-tail-conditions} at \(k_0\) with \(\sigma=1\), so
case~(a) applies. For the zero type, \(k_0=1\) and
\eqref{eq:completion-matching} reads \(0\ge3(1-q)/8\).
\end{proof}

\subsection{Proof of \texorpdfstring{\cref{lem:family-matching}}{Lemma \getrefnumber{lem:family-matching}}: Pointwise Comparison of Tail Sums}
\label{app:pointwise-comparison}

We prove \cref{lem:family-matching}; recall that \(H_\lambda(q)=(1-\lambda)qp_q+\lambda v(q)\).

\begin{proof}
For the zero type, \(k=1\) and the claim is immediate. Let \(D\) be
positive and put \(\sigma=1-\lambda\). Local optimality is
\eqref{eq:adjacent-tail-conditions}. If \(\sigma=0\), then
\(p_{k+1}=0\), so \(v(k)\ge v(q)\) for every \(q\) and the claim holds
with \(c_1=0\). Otherwise, it is case~(b) of
\cref{lem:completion-comparison}.
\end{proof}

\subsection{Proof of \texorpdfstring{\cref{lem:starting-density}}{Lemma \getrefnumber{lem:starting-density}}: Comparison of Averages}
\label{app:starting-density}

We prove \cref{lem:starting-density}.

\paragraph{Reduction to a single type.}
Put \(\rho=r/u\). If \(\rho\le1/2\), averaging
\eqref{eq:completion-matching} with \(q=Q\) and using
\(u_0\ge u\) and \(3/8\ge3\rho/4\) gives
\[
 r_0\ge\tfrac34r+\tfrac38(u_0-u)
 \ge\tfrac34\rho u+\tfrac34\rho(u_0-u)
 =\tfrac34\rho u_0\ge\alpha\rho u_0.
\]
If \(\rho>1/2\), it suffices to find a weight \(\kappa(t)\ge0\),
common to all log-concave types, such that every type, every
\(q\in\mathcal Q\), and every \(t\in[1/2,1)\) satisfy
\begin{equation}
 v(k_0)-\alpha tk_0\ge\kappa(t)\bigl(v(q)-tq\bigr),
 \qquad k_0=k_0(D).
 \label{eq:family-density}
\end{equation}
Indeed, averaging with \(q=Q\) at any \(t\in[1/2,\rho)\) gives
\(r_0-\alpha tu_0\ge\kappa(t)(r-tu)\ge0\), and letting \(t\uparrow\rho\)
proves the claim, including when \(\rho=1\). This case does not use
\(u\le u_0\).

We prove \eqref{eq:family-density} in three steps:
\cref{lem:surplus-extension} reduces it to the single comparison
\(q=1\); \cref{lem:prefix-envelopes} bounds the two quantities this
comparison involves; and \cref{lem:endpoint-checks} verifies the
resulting scalar inequalities. The weight \(\kappa\) is only a proof
device. We say that \(k\) satisfies the \emph{prefix condition} if
\begin{equation}
 k\ge2,\qquad p_k>0,\qquad p_k\ge(k-1)d_{k-1}.
 \label{eq:prefix-condition}
\end{equation}
If \(k_0(D)\ge2\), then \(k_0(D)\) satisfies it, since it maximises
\(jp_j\). Let
\[
 L_\alpha(t):=\frac{2t(1-\alpha t)}{1-t^2}.
\]

\begin{lemma}[Reduction to the first partial sum]
\label{lem:surplus-extension}
Let \(\kappa:[1/2,1)\to\mathbb R\) be nondecreasing with
\(\alpha\le\kappa(t)\le L_\alpha(t)\). Suppose that every positive
log-concave type, every \(t\in[1/2,1)\), and every \(k\) satisfying
\eqref{eq:prefix-condition} satisfy
\begin{equation}
 v(k)-\alpha tk\ge\kappa(t)(p_1-t).
 \label{eq:first-sum-comparison}
\end{equation}
Then \eqref{eq:family-density} holds for every log-concave type.
\end{lemma}

\begin{proof}
For the zero type, \(k_0=1\), and \eqref{eq:family-density} follows
from \(\kappa\ge\alpha\). Let \(D\) be positive, write \(k=k_0(D)\)
and \(\kappa=\kappa(t)\), and recall that \(k\) satisfies
\eqref{eq:adjacent-tail-conditions} with \(\sigma=1\).

\emph{Quantities at or above \(k\).}
Put \(q=k+\ell\) and \(x=\ell/k\). By \cref{lem:tail-estimates}(ii)
and \(p_k\le1\),
\[
 \bigl(v(q)-tq\bigr)-\bigl(v(k)-tk\bigr)
 \le kp_k(1-e^{-x}-tx)
 \le v(k)\frac{(1-t)^2}{2t},
\]
because \(kp_k\le v(k)\) and
\(\sup_{x\ge0}(1-e^{-x}-tx)=1-t+t\log t
=\int_t^1\frac{y-t}{y}\,dy\le\frac{(1-t)^2}{2t}\).
Thus the gap in \eqref{eq:family-density} is at least
\[
 \left(1-\kappa\frac{1+t^2}{2t}\right)v(k)+(\kappa-\alpha)tk.
\]
This is affine in \(v(k)\in[0,k]\), and it is nonnegative at
\(v(k)=0\) because \(\kappa\ge\alpha\) and at \(v(k)=k\) because
\(\kappa\le L_\alpha(t)\).

\emph{Quantities below \(k\).}
It suffices to take \(q\) to be the smallest maximiser of
\(v(q)-tq\). If \(q\ge k\), the previous case applies; if \(q=1<k\),
the claim is \eqref{eq:first-sum-comparison}. Otherwise
\(1<q<k\); put \(a=q-1\) and \(w=p_a\). Minimality of \(q\) gives
\(p_q>t\), so \(w>t\). Conditioning on \(R\ge a\) and subtracting
\(a\) gives a positive log-concave type with tails
\(p'_j=p_{a+j}/w\), and \(k'=k-a\ge2\) satisfies
\eqref{eq:prefix-condition} for it, because \(k\) does. Since
\(t/w\in[1/2,1)\), applying \eqref{eq:first-sum-comparison} at
\(t/w\) and multiplying by \(w\) gives
\[
 v(k)-v(a)-\alpha t(k-a)\ge\kappa(t/w)(p_q-t)\ge\kappa(p_q-t),
\]
using that \(\kappa\) is nondecreasing and \(p_q>t\). Moreover,
\(ta\le v(a)\le a\), and
\(v(a)-\alpha ta\ge\kappa(v(a)-ta)\): this affine inequality in
\(v(a)\) holds at \(v(a)=ta\) because \(\alpha\le1\), and at
\(v(a)=a\) because \(\kappa\le L_\alpha(t)\le(1-\alpha t)/(1-t)\).
Adding the two inequalities gives \eqref{eq:family-density}.
\end{proof}

\begin{lemma}[Prefix bounds]
\label{lem:prefix-envelopes}
\label{lem:lc-prefix}
If \(k\) satisfies \eqref{eq:prefix-condition} for a positive
log-concave type, then
\[
 p_1\ge1-\frac1k\qquad\text{and}\qquad v(k)\ge p_1(\alpha k+\beta_e).
\]
\end{lemma}

\begin{proof}
Put \(w=1-1/k\). Substituting \(d_{k-1}=p_{k-1}-p_k\) into
\(p_k\ge(k-1)d_{k-1}\) gives \(p_k/p_{k-1}\ge w\). Tail ratios are
nonincreasing, so \(p_j/p_{j-1}\ge w\) for \(1\le j\le k\). Hence
\(p_1\ge w\) and \(p_j\ge p_1w^{j-1}\), and summing gives
\(v(k)\ge p_1k(1-w^k)\).

It remains to show \(w^k\le e^{-1}(1-1/(2k))\), since then
\(k(1-w^k)\ge\alpha k+\beta_e\). Put \(x=1/k\le1/2\). The series of the
logarithm gives \(\log(1-x)\le-x-x^2/2-x^3/3\) and, for
\(0\le u\le1/2\), \(\log(1-u)\ge-u-u^2\). Therefore
\[
 \log w^k=\frac{\log(1-x)}x\le-1-\frac x2-\frac{x^2}3
 \le-1-\frac x2-\frac{x^2}4\le-1+\log\Bigl(1-\frac x2\Bigr).
 \qedhere
\]
\end{proof}

By \cref{lem:prefix-envelopes}, the gap in
\eqref{eq:first-sum-comparison} is at least
\[
 E_k(p_1;t),\qquad
 E_k(x;t):=x(\alpha k+\beta_e)-\alpha tk-\kappa(t)(x-t),
\]
which is affine in \(x=p_1\in[1-1/k,1]\). It therefore remains to
choose \(\kappa\) and check the two endpoints. Throughout, put
\(\delta=1-t\in(0,1/2]\).

\begin{lemma}[Endpoint checks]
\label{lem:endpoint-checks}
\label{lem:lc-density}
The weight
\[
 \kappa(t):=\beta_e\Bigl(2+\frac1\delta\Bigr)
\]
is nondecreasing in \(t\), satisfies
\(\alpha\le\kappa(t)\le L_\alpha(t)\), and gives \(E_k(1;t)\ge0\) and
\(E_k(1-1/k;t)\ge0\) for every \(k\ge2\) and \(t\in[1/2,1)\).
\end{lemma}

\begin{proof}
\emph{The weight.} It is increasing in \(t\), and
\(\kappa\ge4\beta_e>\alpha\). Since \(1-\alpha t=2\beta_e+\alpha\delta\)
and \(\alpha=1-2\beta_e\ge3\beta_e\),
\[
 L_\alpha(t)=\frac{2t(2\beta_e+\alpha\delta)}{\delta(1+t)}
 \ge\frac{2\beta_e\,(1-\delta)(2+3\delta)}{\delta(2-\delta)},
 \qquad
 \kappa(t)=\frac{\beta_e(1+2\delta)}{\delta}.
\]
So \(\kappa\le L_\alpha\) follows from
\(2(1-\delta)(2+3\delta)-(1+2\delta)(2-\delta)
=2-\delta-4\delta^2>0\).

\emph{The endpoint \(x=1\).} Using \(\kappa\delta=\beta_e+2\beta_e\delta\)
and \(k\ge2\),
\[
 E_k(1;t)=\beta_e+\alpha k\delta-\kappa\delta
 =(\alpha k-2\beta_e)\delta>0.
\]

\emph{The endpoint \(x=1-1/k\).} Using in addition
\(\kappa-\beta_e=\beta_e(1+\delta)/\delta\) and the AM--GM inequality,
\[
 E_k\Bigl(1-\frac1k;t\Bigr)
 =\alpha k\delta+\frac{\beta_e(1+\delta)}{\delta k}-\alpha-2\beta_e\delta
 \ge2\sqrt{\alpha\beta_e(1+\delta)}-(\alpha+2\beta_e\delta).
\]
This is nonnegative because
\(4\alpha\beta_e(1+\delta)-(\alpha+2\beta_e\delta)^2
=\alpha(4\beta_e-\alpha)-4\beta_e^2\delta^2
\ge\alpha(4\beta_e-\alpha)-\beta_e^2>0\); numerically,
\(\alpha(4\beta_e-\alpha)\approx0.066\) and \(\beta_e^2\approx0.034\).
\end{proof}

\paragraph{Completing the proof.}
\Cref{lem:prefix-envelopes,lem:endpoint-checks} give
\eqref{eq:first-sum-comparison} with \(\kappa(t)=\beta_e(2+1/\delta)\),
and this weight satisfies the hypotheses of
\cref{lem:surplus-extension}. Hence \eqref{eq:family-density} holds,
which completes the proof of \cref{lem:starting-density}.

\Needspace{20\baselineskip}
\section{Omitted Proofs for \texorpdfstring{\cref{sec:restricted}}{Section \getrefnumber{sec:restricted}}: Impossibility Bound}
\label{app:upper-bounds}

The upper bound follows from a weighted sum of incentive constraints
between neighbouring types. We first prove this argument in general and
then apply it to truncated geometric types.

\subsection{Neighbouring-Type Construction}
\label{sec:common-interval-upper}
\label{app:shared-upper}

\begin{lemma}[Neighbouring-type construction]
\label{lem:neighbouring-upper}
Fix integers \(M>L\ge3\) and \(m\ge M\). Let \(D_b\) be a type for
\(b=L-1,\ldots,M\), and write
\(p_b(z)=\Pr_{D_b}(R\ge z)\), including \(p_b(0)=1\),
and \(v_b(q)=\mathbb E_{D_b}[\min\{R,q\}]\).
For \(K,s>0\), put
\[
 g_b(z)=K\bigl[b p_b(z)-(b+1)p_{b-1}(z)\bigr]
 \qquad(L\le b\le M).
\]
Suppose for these indices that
\begin{equation}
 p_b(1)+g_b(0)\le s,\qquad
 \frac{v_b(z)}z+g_b(z)\le s\quad(1\le z\le m).
 \label{eq:neighbouring-test}
\end{equation}
Give the types probabilities \(w_b/W\), where
\begin{equation}
 w_{L-1}=\frac{K}{sL},\qquad
 w_b=\frac1{b(b+1)}\ (b\ge L),\qquad
 W=\sum_{b=L-1}^M w_b.
 \label{eq:common-upper-population}
\end{equation}
If \(0<\xi<1/2\), \(L\ge2/\xi\), and
\[
 p_{L-1}(1)\ge\rho>0,\qquad
 p_b(\lfloor\xi b\rfloor)\ge\rho\quad(b\ge L),
\]
then every admissible fixed menu satisfies
\begin{equation}
 \limsup_{n\to\infty}
 \frac{\operatorname{Rec}(\mathcal M;B,B)}{\mathrm{OPT}(B)}
 \le\frac{s+\Gamma/\Lambda}{\rho},
 \quad
 \Gamma=w_{L-1}\mathbb E_{D_{L-1}}R,\qquad
 \Lambda=w_{L-1}+\sum_{b=L}^M w_b\lfloor\xi b\rfloor.
 \label{eq:common-upper-ratio}
\end{equation}
Here \(B=\lfloor n\Lambda/W\rfloor\), with the population and \(m\)
fixed as \(n\to\infty\).
\end{lemma}

\begin{proof}
Because \(v_b(q)/q\) is nonincreasing,
\eqref{eq:neighbouring-test} implies, for every \(0\le z\le q\),
\begin{equation}
 g_b(z)\le s-\frac{v_b(q)}q.
 \label{eq:diagonal-extension}
\end{equation}
Let \(C_b=(q_b,\tau_b)\) be a contract selected by type \(b\), and
write \(\tau_b(z)=\sum_{j=0}^{q_b}a_{bj}\mathbf1\{z\ge j\}\) with
\(a_{bj}\ge0\). Then \(U_{b'}(C_b)=\sum_ja_{bj}p_{b'}(j)\) and
\(P_b=\sum_ja_{bj}\ge q_b\). Since each type weakly prefers its own
contract to its neighbour's,
\[
 \Delta:=\sum_{b=L-1}^{M-1}\frac K{b+1}
 \bigl(U_b(C_b)-U_b(C_{b+1})\bigr)\ge0.
\]
In \(\Delta\), the coefficient of \(a_{bj}\) is
\(Kp_b(j)/(b+1)-Kp_{b-1}(j)/b=w_bg_b(j)\) for \(L\le b<M\). For
\(b=M\), only the second term appears, so the coefficient is at most
\(w_Mg_M(j)\); for \(b=L-1\), only the first appears, giving
\((K/L)p_{L-1}(j)\le sw_{L-1}\). By \eqref{eq:diagonal-extension}, each
type \(b\ge L\) therefore contributes at most
\[
 w_bv_b(q_b)-sw_bP_b+w_bP_b\left(s-\frac{v_b(q_b)}{q_b}\right)
 =w_bv_b(q_b)\left(1-\frac{P_b}{q_b}\right)\le0
\]
to \(\sum_bw_b(v_b(q_b)-sP_b)+\Delta\), and type \(L-1\) contributes
at most \(w_{L-1}v_{L-1}(q_{L-1})\le\Gamma\). Averaging over tied
choices and using \(\Delta\ge0\) gives \(r\le se+\Gamma/W\). Hence
\[
 \operatorname{Rec}(\mathcal M;B,B)
 \le r\min\{n,B/e\}\le sB+(\Gamma/W)n.
\]

The oracle assigns one voucher to type \(L-1\) and
\(\lfloor\xi b\rfloor\ge1\) vouchers to each type \(b\ge L\). Its mean
quantity is \(\Lambda/W\), and the tail assumptions give mean recruitment at
least \(\rho\Lambda/W\). Since \(B=\lfloor n\Lambda/W\rfloor\),
\cref{thm:relaxation-transfer} shows that it recruits at least
\(\rho\Lambda n/W-o(n)\). Dividing the menu bound by this quantity proves
\eqref{eq:common-upper-ratio}.
\end{proof}

When \(\mathbb E_{D_{L-1}}R\le L\), the boundary cost satisfies
\(\Gamma\le K/s\), whereas
\[
 \Lambda\ge\frac\xi2\sum_{b=L}^M\frac1{b+1}\longrightarrow\infty
 \qquad(M\to\infty).
\]
Thus, when \(\rho\to1\) as \(\xi\to0\), the ratio approaches
\(s\). For a prescribed error, choose \(\xi\), then \(L\), then
a finite \(M\); fix this population before taking the budget limit.

\Needspace{13\baselineskip}
\subsection{Proof of \texorpdfstring{\cref{thm:uniform-upper}}{Theorem \getrefnumber{thm:uniform-upper}}: Log-Concave Types}
\label{app:uniform-upper}
\label{app:linear-upper}
\label{app:lc-upper}

We apply \cref{lem:neighbouring-upper} to truncated geometric types,
with \(K:=e-2\) and \(\alpha:=1-1/e\).

\begin{lemma}[A scalar inequality]
\label{lem:lc-scalar}
For every \(y>0\),
\(\displaystyle
 H(y):=\frac{1-e^{-y}}y-K(1-y)e^{-y}\le1-\frac1e.
\)
\end{lemma}

\begin{proof}
Since \((e^y-1)/y=\int_0^1e^{ys}\,ds\),
\[
 \varphi(y):=e^{y}\bigl(\alpha-H(y)\bigr)
 =\alpha e^{y}-\int_0^1e^{ys}\,ds+K(1-y).
\]
Then
\(\varphi''(y)=\alpha e^y-\int_0^1s^2e^{ys}\,ds
\ge(\alpha-\tfrac13)e^y>0\), so \(\varphi\) is convex. Moreover
\(\varphi(1)=\alpha e-(e-1)=0\) and
\(\varphi'(1)=\alpha e-\int_0^1se^{s}\,ds-K=(e-1)-1-(e-2)=0\).
A convex function lies above its tangent, so \(\varphi\ge0\).
\end{proof}

\begin{proof}[Proof of \cref{thm:uniform-upper}]
Fix \(\zeta\in(0,1/2)\). We choose, in this order, \(\xi\le\zeta\),
integers \(M>L\ge\max\{3,2/\xi\}\), and \(m\ge M\); the last three are
specified below. For \(L-1\le b\le M\), let \(D_b\) be the geometric
type with ratio \(e^{-1/b}\) conditioned on \(R\le m\). Its masses are
proportional to \(e^{-j/b}\) on \(\{0,\ldots,m\}\), so it is
log-concave, and
\[
 p_b(z)=\frac{e^{-z/b}-\epsilon_b}{1-\epsilon_b},
 \qquad \epsilon_b:=e^{-(m+1)/b}.
\]
Let \(\bar\epsilon:=\epsilon_M/(1-\epsilon_M)\), which tends to zero
as \(m\to\infty\). Then
\begin{equation}
 e^{-z/b}-\bar\epsilon\le p_b(z)\le e^{-z/b}
 \qquad(0\le z\le m).
 \label{eq:lc-truncation}
\end{equation}

\emph{The test \eqref{eq:neighbouring-test}.} Let \(L\le b\le M\),
\(1\le z\le m\), and \(y=z/b\). By \eqref{eq:lc-truncation},
\[
 \frac{v_b(z)}z\le\frac1z\int_0^{z}e^{-x/b}\,dx=\frac{1-e^{-y}}y,
 \qquad
 g_b(z)\le K\bigl[b\,e^{-z/b}-(b+1)e^{-z/(b-1)}\bigr]
          +K(b+1)\bar\epsilon.
\]
Since \(z/(b-1)=y+y/(b-1)\) and \(e^{-x}\ge1-x\),
\[
 b\,e^{-z/b}-(b+1)e^{-z/(b-1)}
 \le e^{-y}\Bigl[b-(b+1)\Bigl(1-\frac y{b-1}\Bigr)\Bigr]
 =-(1-y)e^{-y}+\frac{2ye^{-y}}{b-1}
 \le-(1-y)e^{-y}+\frac{2}{e(L-1)}.
\]
With \cref{lem:lc-scalar}, this gives
\[
 \frac{v_b(z)}z+g_b(z)
 \le H(y)+\frac{2K}{e(L-1)}+K(M+1)\bar\epsilon
 \le\alpha+2\zeta=:s,
\]
once \(L\ge1+2K/(e\zeta)\) and \(m\) is large enough that
\(K(M+1)\bar\epsilon\le\zeta\). At \(z=0\),
\(p_b(1)+g_b(0)\le1-K<\alpha\le s\).

\emph{The oracle.} By \eqref{eq:lc-truncation} and
\(L-1\ge1/\xi\), the tails \(p_b(\lfloor\xi b\rfloor)\) and
\(p_{L-1}(1)\) are at least
\(\rho:=e^{-\xi}-\bar\epsilon\ge e^{-\zeta}-\zeta>0\). Moreover
\[
 \Gamma\le\frac K{sL}\sum_{z\ge1}e^{-z/(L-1)}\le\frac Ks,
 \qquad
 \Lambda\ge\frac\xi2\sum_{b=L}^{M}\frac1{b+1}\longrightarrow\infty
 \quad(M\to\infty),
\]
so we may choose \(M\) with \(\Gamma/\Lambda\le\zeta\).
\Cref{lem:neighbouring-upper} then gives, for every admissible fixed
menu,
\[
 \limsup_{n\to\infty}
 \frac{\operatorname{Rec}(\mathcal M;B,B)}{\mathrm{OPT}(B)}
 \le\frac{s+\Gamma/\Lambda}{\rho}
 \le\frac{1-1/e+3\zeta}{e^{-\zeta}-\zeta}.
\]
The right-hand side tends to \(1-1/e\) as \(\zeta\to0\), so it is at
most \(1-1/e+\eta\) for \(\zeta\) small enough. Fixing the resulting
population and \(m\) before taking the budget limit proves the claim.
\end{proof}

\section{Omitted Details for \texorpdfstring{\cref{sec:experiments}}{Section \getrefnumber{sec:experiments}}: Experiments}
\label{app:experiments}

\subsection{Dataset and Prior Construction}
\label{app:experiments-data}

\paragraph{Network.}
We use the Persons, Egodyads, and Altdyads files of ICPSR study 22140
\citep{morris2011hiv} and reconstruct the procedure of
\citet{pan2026adaptive}. We keep relationships whose endpoints both
have known HIV status and form a simple undirected graph whose nodes
are (study, person) pairs, so that identical identifiers in
different studies remain distinct. Repeated edges are merged, and
only endpoints of retained edges enter the graph. The graph has
\(1{,}386\) nodes from five studies and \(3{,}796\) edges; degrees
range from \(1\) to \(74\), with mean \(5.48\). As in the reference
implementation, the three self-loops are retained, and a person with
inconsistent recorded statuses takes the maximum.

\paragraph{Types of \citet{pan2026adaptive}.}
A decision-tree classifier predicts each node's degree from
\(17\) categorical fields of the Persons file
(\texttt{LOCAL}, \texttt{RACE}, \texttt{ETHN}, \texttt{SEX},
\texttt{ORIENT}, \texttt{BEHAV}, \texttt{PRO}, \texttt{PIMP},
\texttt{JOHN}, \texttt{DEALER}, \texttt{DRUGMAN}, \texttt{THIEF},
\texttt{RETIRED}, \texttt{HWIFE}, \texttt{DISABLE}, \texttt{UNEMP},
\texttt{STREETS}), one-hot encoded with missing-value codes as
categories. As in the reference implementation, the features also
incorporate each person's HIV status. The tree has minimum leaf size
\(\lfloor0.01\cdot1{,}386\rfloor=13\) and no depth limit, and
produces \(64\) leaves of \(13\) to \(40\) people. Leaf \(j\), with
\(N_j\) people, defines type \(j\) with share \(w_j=N_j/1{,}386\) and
degree distribution \(H_j\), the empirical degree distribution of its
members. The prior of \citet{pan2026adaptive} uses \(H_j\) as the
capacity distribution of type \(j\). Since nodes enter the graph only
through retained edges, every degree is at least one, so in this
prior no type ever recruits nobody. The reference implementation does
not fix the tree's random seed, so our tree, with a fixed seed,
reconstructs its procedure rather than reproducing a saved tree.

\paragraph{Our modification.}
We draw a degree \(d\sim H_j\), convert each contact independently
with probability \(0.7\), and cap the result at \(m=20\):
\(R=\min\{X,20\}\) with \(X\mid d\sim\mathrm{Bin}(d,0.7)\). Thus,
for \(r<20\),
\[
 D_j(r)=\sum_{d\ge r}H_j(d)\binom dr0.7^r\,0.3^{d-r},
 \qquad
 D_j(20)=\sum_dH_j(d)\Pr\bigl[\mathrm{Bin}(d,0.7)\ge20\bigr].
\]
Conversion is applied before the cap. It gives every type a positive
probability of recruiting nobody, reflecting that not every contact
is successfully recruited; the probability \(0.7\) is a modelling
assumption, not an estimate. The cap gives capacities the known bound
\(m\) that our model requires. The prior selects \(D_j\) with
probability \(w_j\). It has mean capacity \(3.67\),
\(\Pr(R=0)=11.2\%\), and \(\Pr(R=20)=2.4\%\), and type means range
from \(0.80\) to \(12.13\).

\paragraph{Scope.}
Simulations draw types independently from this prior; they do not
traverse the network, model overlapping contacts, or let recruits
become referrers. The prior is fitted on the full data without a
held-out split, so the experiments are simulations under an
empirical model rather than a held-out validation. Initial
individuals are not counted in recruitment.

\subsection{Implementation Details}
\label{app:experiments-implementation}

\paragraph{Policies.}
The oracle threshold policy runs \cref{alg:oracle-threshold} with
\(m=20\). Before each individual is served, the remaining budgets
must cover the largest quantity the policy assigns. The constant
allocation offers the single contract \((B_1/n,z\mapsto\min\{z,B_1/n\})\).
The voucher-price menu with cap \(m\) uses the
\((\theta^\star,\eta^\star)\) computed for \(m=20\), allocates
\(\min\{A^\star(D),m\}\), and reserves \(q+\theta^\star(m-q)\) for
quantity \(q\). The best fitted cap maximises, over
\(m\in\{1,\ldots,20\}\), the relaxed recruitment
\(r_m\min\{n,B_1/u_m,B_2/e_m\}\) under the fitted prior, where
\(u_m\) and \(r_m\) are the mean allocation and recruitment of the
capped menu and \(e_m=u_m+\theta^\star(m-u_m)\) is its mean
reservation. The reduced cap of \cref{sec:exp-misspecified} is
\(\max\{1,m^\star-3\}\), where \(m^\star\) is the best fitted cap; it
keeps the same fitted cutoff and tie rule and is not reselected using
the true prior. The completion-bonus menu runs
\cref{alg:restricted-menu} with \(m=20\), including its all-or-nothing
case. For every menu, the stopping rule of
\cref{sec:recruitment-process} requires the remaining budgets to
cover every offered contract, and unused reservations are not
recycled within the round.

\paragraph{Parameter tuning.}
For \cref{sec:exp-tuning}, we fix the best fitted cap and the tie
probability of the algorithm, and vary \(\theta\) on a grid of
spacing \(0.005\) on \([0,1]\), refined locally around the five best
grid points and at the values where individuals' choices change. We
tune \(\lambda\) for the completion bonus in the same way, with
\(m=20\). The tuned parameter maximises mean recruitment over \(100\)
tuning trials and is then evaluated on \(100\) independent trials.
Tuning is conditional on the selected cap, not a joint optimisation
over caps and prices. The reported gain is the relative change in mean
recruitment, with a paired delta-method \(95\%\) confidence interval.

\paragraph{Normalisation and error bars.}
The denominator \(F(B)\) is computed on the true population. For the
base prior, \(F(B)=2{,}722.7\), \(3{,}031.0\), and \(3{,}255.9\) at
\(B_1/n=4,5,6\); these values do not depend on \(B_2\), since
\(B_2\ge B_1\) in every setting. For the experiment of
\cref{sec:exp-extreme}, \(F(B)\) is recomputed on each modified
population. In the experiment of \cref{sec:exp-misspecified}, the same
\(100\) draws of \(Z\) are reused across tilt strengths, methods, and
budgets, and the denominator is the true \(F(B)\). Each plotted point
is \(\overline R/F(B)\pm1.96\,s/(10F(B))\), where \(\overline R\) and
\(s\) are the sample mean and standard deviation of recruitment over
the \(100\) trials. Each experiment draws its own trials, so the same
setting can differ in the second decimal place across figures. These
intervals reflect simulation randomness only, not uncertainty in
estimating the prior from the data.

\subsection{Additional Budgets}
\label{app:experiments-additional}

\begin{figure}[!htbp]
\centering
\includegraphics[width=0.8\textwidth]{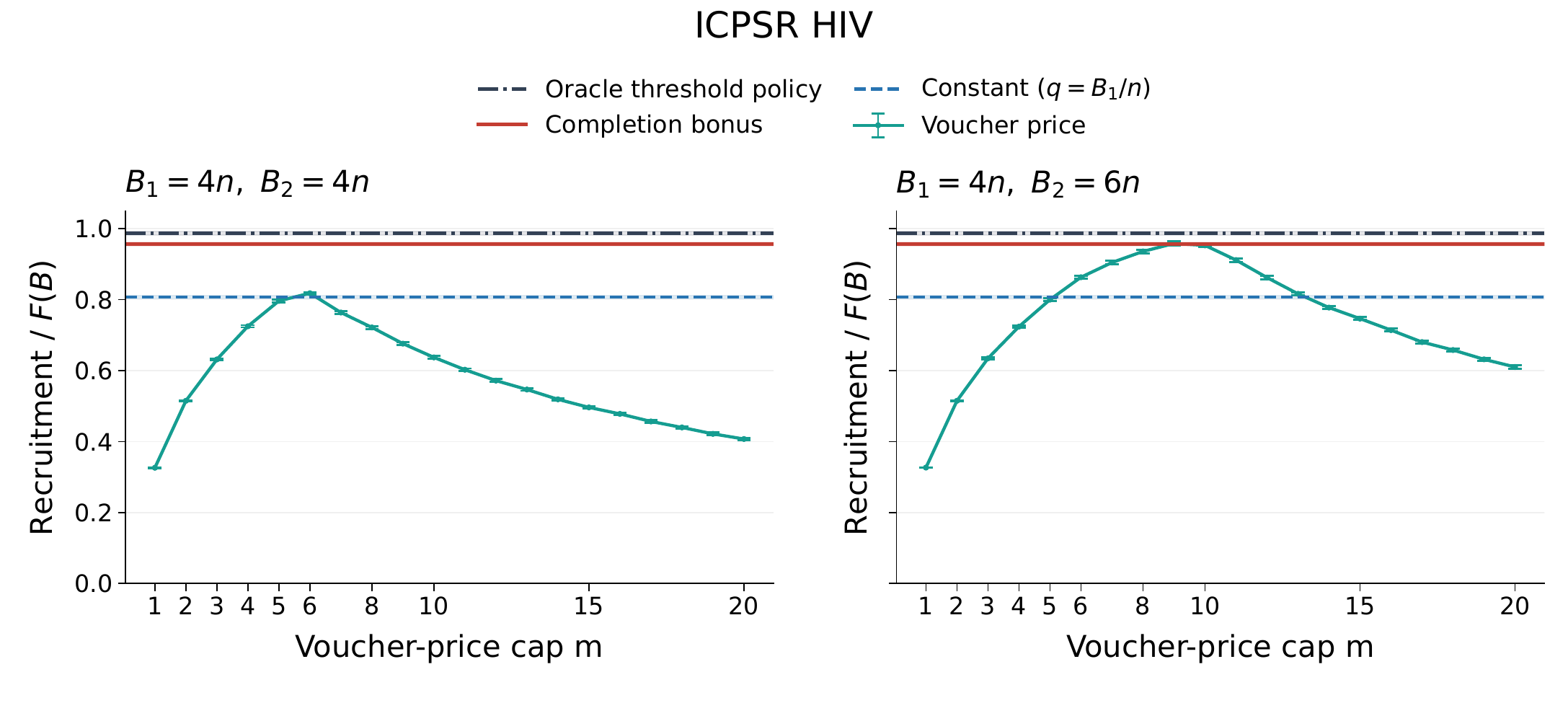}\\[4pt]
\includegraphics[width=0.8\textwidth]{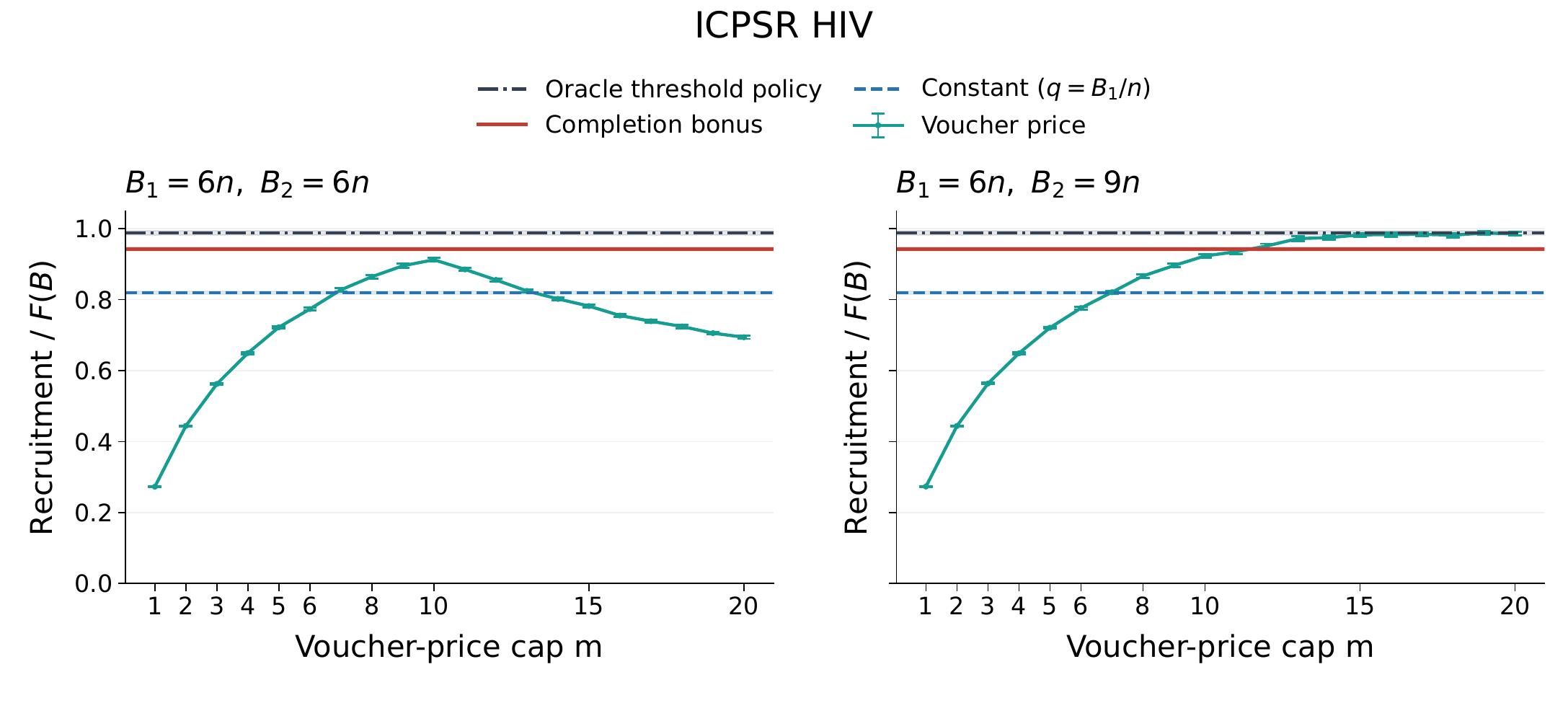}
\caption{Effect of the voucher cap, as in \cref{fig:exp-caps}, with
\(B_1=4n\) (top) and \(B_1=6n\) (bottom), and \(B_2=B_1\) (left) or
\(B_2=1.5B_1\) (right).}
\label{fig:exp-caps-app}
\end{figure}

\begin{figure}[!htbp]
\centering
\includegraphics[width=0.8\textwidth]{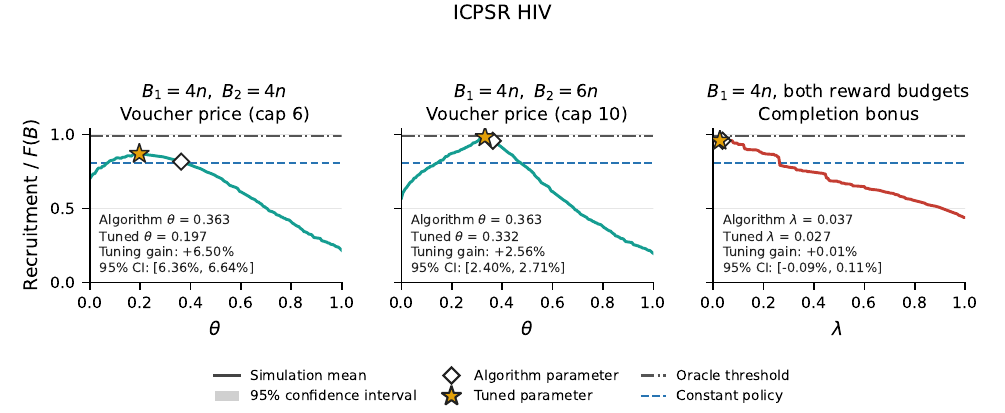}\\[4pt]
\includegraphics[width=0.8\textwidth]{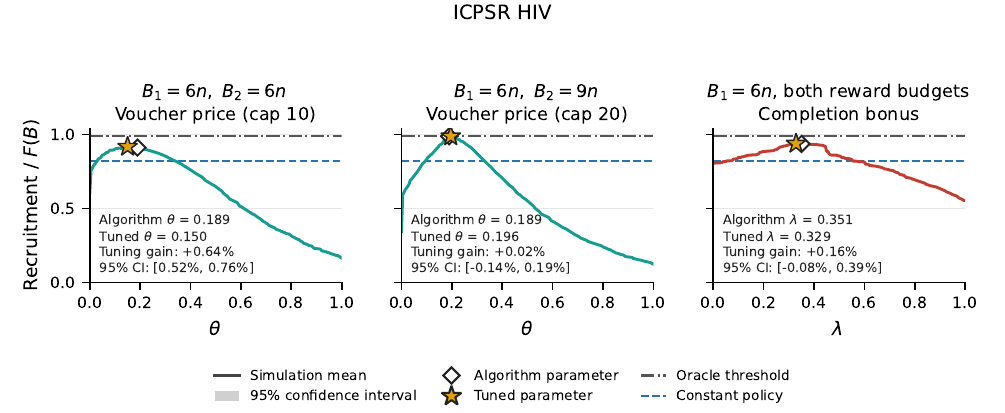}
\caption{Distance from optimal parameters, as in
\cref{fig:exp-tuning}, with \(B_1=4n\) (top) and \(B_1=6n\)
(bottom).}
\label{fig:exp-tuning-app}
\end{figure}

\begin{figure}[!htbp]
\centering
\includegraphics[width=0.8\textwidth]{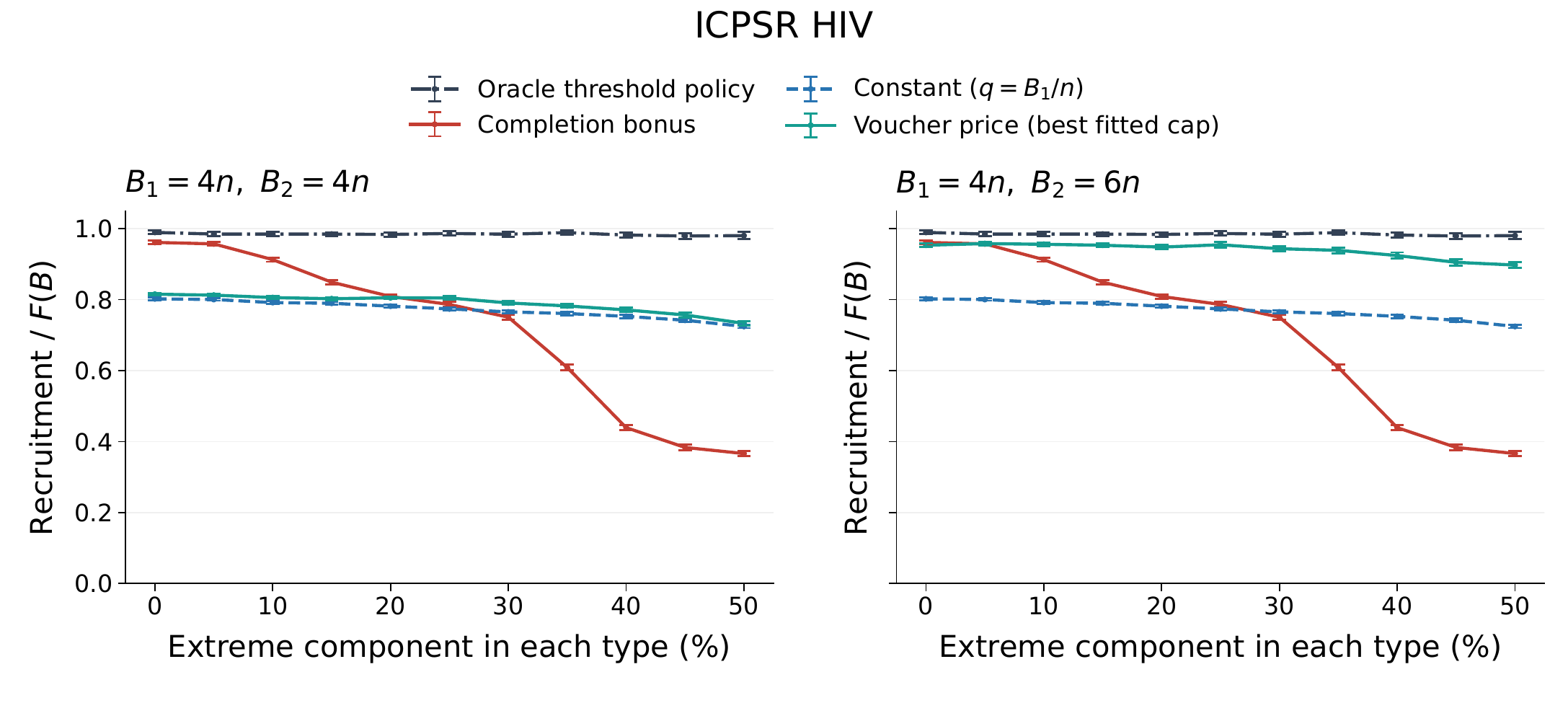}\\[4pt]
\includegraphics[width=0.8\textwidth]{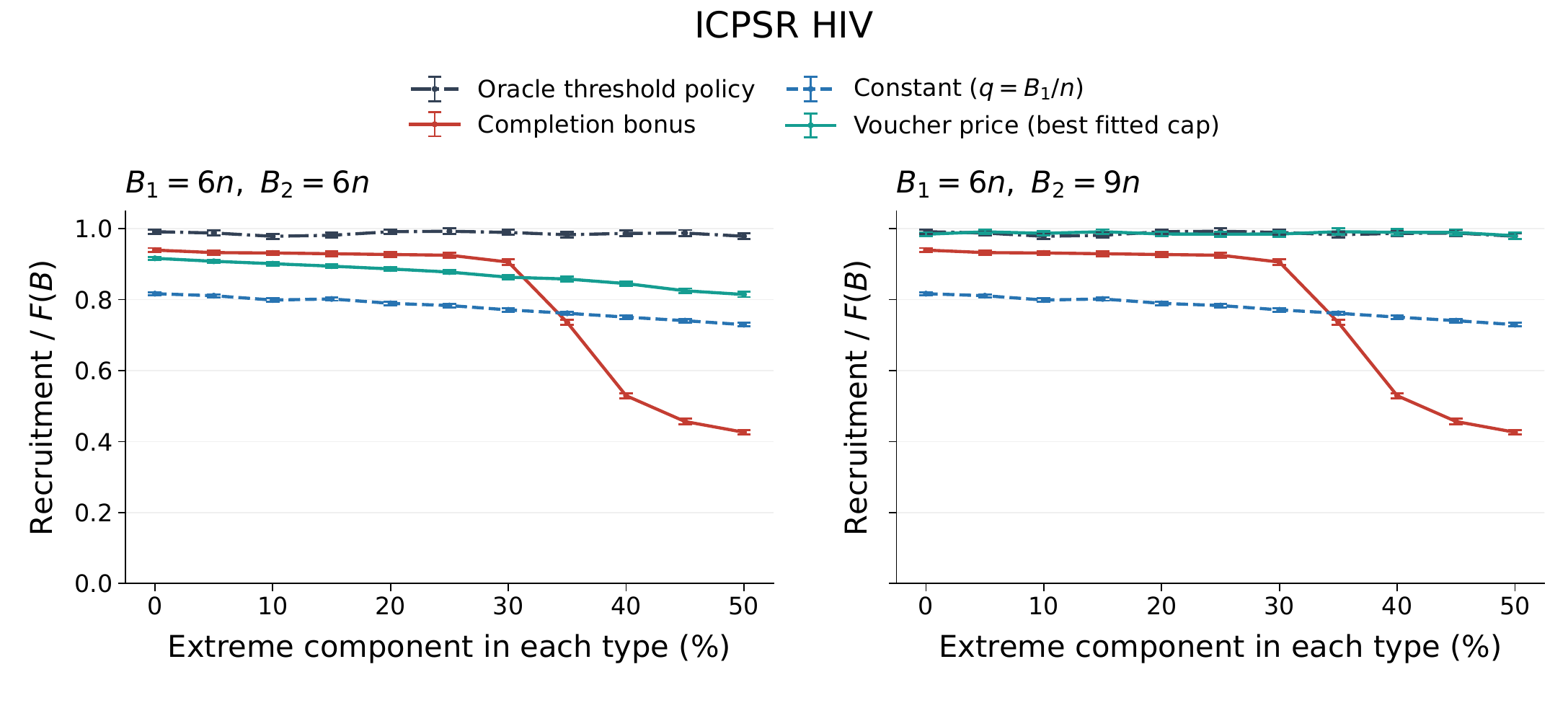}
\caption{More extreme types, as in \cref{fig:exp-extreme}, with
\(B_1=4n\) (top) and \(B_1=6n\) (bottom), and \(B_2=B_1\) (left) or
\(B_2=1.5B_1\) (right).}
\label{fig:exp-extreme-app}
\end{figure}

\begin{figure}[!htbp]
\centering
\includegraphics[width=0.8\textwidth]{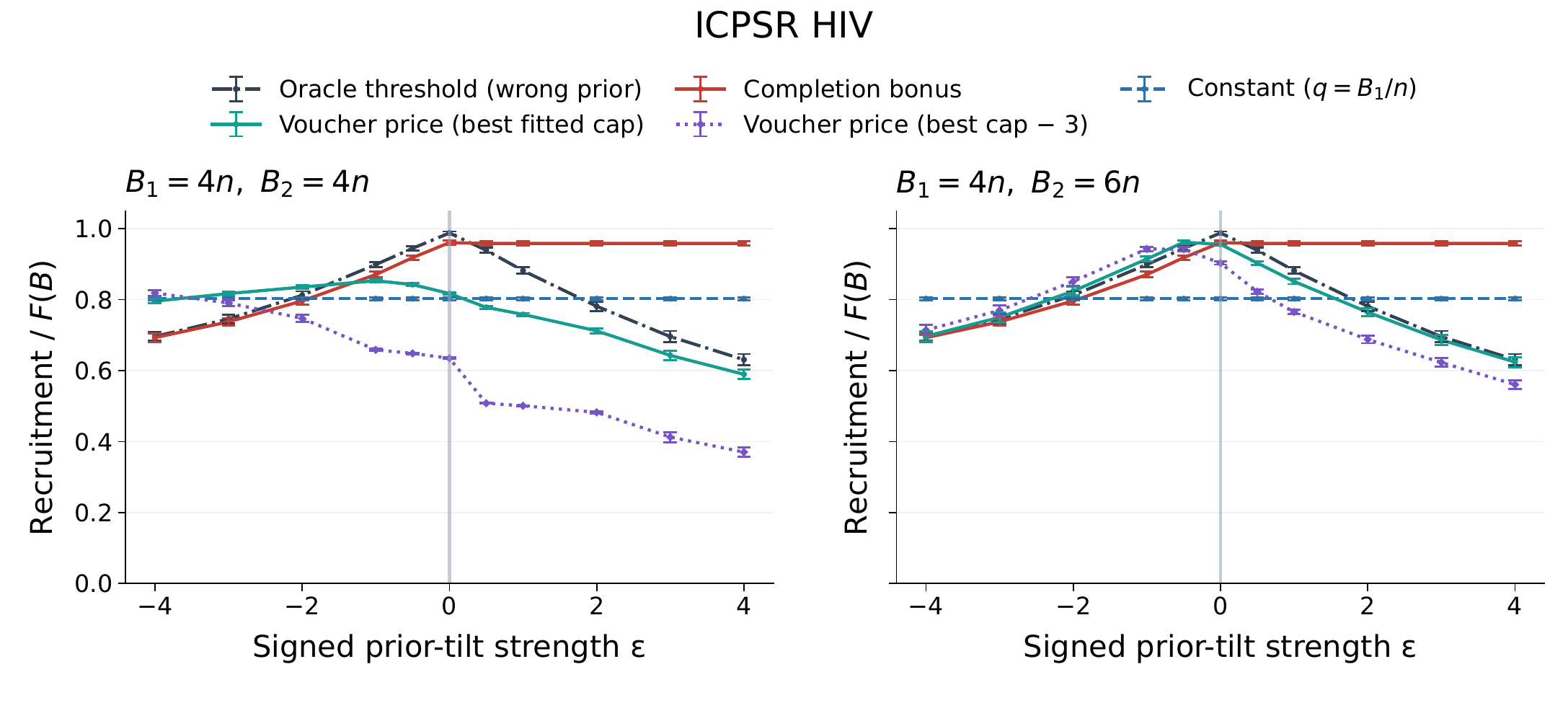}\\[4pt]
\includegraphics[width=0.8\textwidth]{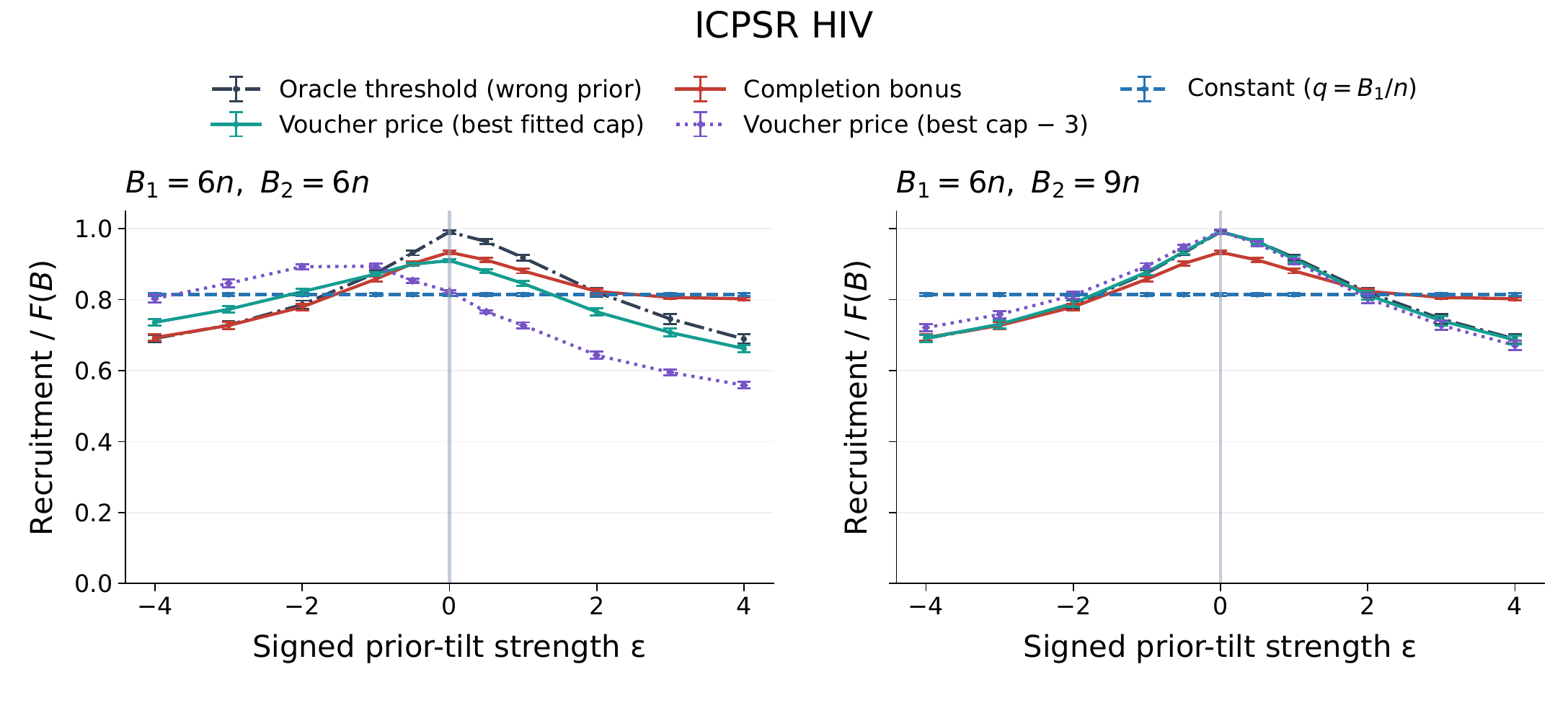}
\caption{Misspecified type shares, as in \cref{fig:exp-prior}, with
\(B_1=4n\) (top) and \(B_1=6n\) (bottom), and \(B_2=B_1\) (left) or
\(B_2=1.5B_1\) (right).}
\label{fig:exp-prior-app}
\end{figure}

\Cref{fig:exp-caps-app,fig:exp-tuning-app,fig:exp-extreme-app,fig:exp-prior-app}
repeat the experiments of \cref{sec:experiments} with \(B_1=4n\) and
\(B_1=6n\), each with \(B_2=B_1\) and \(B_2=1.5B_1\).

\paragraph{Voucher caps (\cref{fig:exp-caps-app}).}
The best cap grows with the budget and with \(B_2/B_1\). In the cap
sweep, the best \emph{observed} caps are, for \(B_2=B_1\) and
\(B_2=1.5B_1\) respectively, \(m=6\) and \(m=9\) at \(B_1=4n\), and
\(m=10\) and \(m=19\) at \(B_1=6n\). The best \emph{fitted} caps,
which maximise relaxed recruitment under the prior and are used in
\cref{fig:exp-tuning-app,fig:exp-extreme-app,fig:exp-prior-app},
coincide with these when \(B_2=B_1\), but are \(m=10\) at
\(B_1=4n\) and \(m=20\) at \(B_1=6n\) when \(B_2=1.5B_1\), within
\(0.5\) and \(0.2\) points of \(F(B)\) of the best observed caps. (At
\(B_1=5n\), the two coincide.) When \(B_2=B_1\), the completion bonus
outperforms every cap (\(0.96\) versus \(0.82\) at \(B_1=4n\), and
\(0.94\) versus \(0.91\) at \(B_1=6n\)). When \(B_2=1.5B_1\), the best
observed cap matches the completion bonus at \(B_1=4n\) (both
\(0.96\)) and outperforms it at \(B_1=6n\) (\(0.99\) versus
\(0.94\)). The constant allocation attains \(0.81\)
and \(0.82\); at \(B_1=4n\) and \(B_2=B_1\), the best cap exceeds it
only slightly.

\paragraph{Distance from optimal parameters
(\cref{fig:exp-tuning-app}).}
Tuning changes recruitment under the completion bonus by at most
\(0.16\%\), and no change is statistically significant. For the
voucher price, tuning gains are largest when the reward budget binds
and the voucher budget is small: \(6.5\%\) at \(B_1=4n\) and
\(B_2=B_1\), where \(\theta\) falls from \(0.363\) to \(0.197\), and
\(2.6\%\) at \(B_1=4n\) and \(B_2=1.5B_1\). At \(B_1=6n\), the gains
are \(0.6\%\) and \(0.02\%\).

\paragraph{More extreme types (\cref{fig:exp-extreme-app}).}
As \(\omega\) grows from \(0\) to \(0.5\), the completion bonus loses
between \(0.51\) and \(0.60\) of \(F(B)\), much of it between
\(\omega=0.30\) and \(\omega=0.40\), and falls below the constant
allocation from \(\omega=0.30\) at \(B_1=4n\) and from
\(\omega=0.35\) at \(B_1=6n\). The voucher price with the best fitted
cap declines by at most \(0.10\).

\paragraph{Misspecified priors (\cref{fig:exp-prior-app}).}
The one-sided robustness persists. At \(B_1=4n\), the completion bonus
is essentially flat for \(\varepsilon>0\), at \(0.96\), since its
weight \(\lambda^\star=0.04\) is already close to the all-or-nothing
end; at \(B_1=6n\), it falls from \(0.93\) to \(0.80\) for positive
\(\varepsilon\). When
\(B_2=B_1\), reducing the cap by three again helps under strong
underestimation: at \(\varepsilon=-4\), it raises the voucher price
from \(0.80\) to \(0.82\) at \(B_1=4n\) and from \(0.74\) to \(0.80\)
at \(B_1=6n\), and at \(\varepsilon=-2\) and \(B_1=6n\), from \(0.82\)
to \(0.89\).

\end{document}